\documentclass[preprints,article,accept,moreauthors]{Definitions/mdpi}

\expandafter\let\csname equation*\endcsname\relax
\expandafter\let\csname endequation*\endcsname\relax
\usepackage{amssymb}  % Better maths support & more symbols
\usepackage{bm}  
\usepackage[titletoc,toc,title]{appendix}
\usepackage{subfig}
\usepackage{sidecap}
\usepackage{xcolor}
\usepackage[linesnumbered,ruled,vlined]{algorithm2e}
\definecolor{darkred}{RGB}{135, 33, 9}
\usepackage{xparse}

\newif\iffirstitem
\firstitemtrue

\SetCommentSty{mycommfont}
\newcommand{\nosemic}{\renewcommand{\@endalgocfline}{\relax}}% Drop semi-colon ;
\newcommand{\dosemic}{\renewcommand{\@endalgocfline}{\algocf@endline}}% Reinstate semi-colon ;
\let\oldnl\nl% Store \nl in \oldnl
\newcommand{\nonl}{\renewcommand{\nl}{\let\nl\oldnl}}% Remove line number for one line

\firstpage{1} 
\pubvolume{xx}
\issuenum{1}
\articlenumber{5}
\pubyear{2020}
\copyrightyear{2020}
\history{Received: date; Accepted: date; Published: date}
\Title{Interacting particle solutions of Fokker--Planck equations through gradient--log--density estimation}

\Author{Dimitra Maoutsa $^{1}$\orcidA{}, Sebastian Reich $^{2}$\orcidB{} and Manfred Opper $^{1}$\orcidC{}}

\AuthorNames{Dimitra Maoutsa, Sebastian Reich and Manfred Opper}

\address{%
$^{1}$ \quad Artificial Intelligence Group, Technische Universit\"{a}t Berlin,
Marchstra{\ss}e 23, Berlin 10587, Germany\\
$^{2}$ \quad Institute of Mathematics, University of Potsdam, Karl-Liebknecht-Str. 24/25, 14476 Potsdam, Germany}

\corres{Correspondence: dimitra.maoutsa@tu-berlin.de; manfred.opper@tu-berlin.de }

\abstract{
Fokker--Planck equations are extensively employed in various scientific fields as they characterise the behaviour of stochastic systems at the level of probability density functions. 
Although broadly used, they allow for analytical treatment only in limited settings, and often is inevitable to resort to numerical solutions.
Here, we develop a computational approach for simulating the time evolution of
Fokker—Planck solutions in terms of a mean field limit of an interacting particle system. The interactions between particles 
are determined by the gradient of the logarithm of the particle density, approximated here by a novel statistical estimator. The performance of our method shows promising results, with more accurate and less fluctuating 
statistics compared to direct stochastic simulations of comparable particle number.
Taken together, our framework allows for effortless and reliable particle-based simulations of Fokker--Planck equations in low and moderate dimensions. The proposed gradient--log--density estimator is also of independent interest, for example, in the context of optimal control. 
     }


\begin{document}
%%%%%%%%%%%%%%%%%%%%%%%%%%%%%%%%%%%%%%%%%%

%%%%%%%%%%%%%%%%%%%%%%%%%%%%%%%%%%%%%%%%%%

%The order of the section titles is: Introduction, Materials and Methods, Results, Discussion, Conclusions for these journals: aerospace,algorithms,antibodies,antioxidants,atmosphere,axioms,biomedicines,carbon,crystals,designs,diagnostics,environments,fermentation,fluids,forests,fractalfract,informatics,information,inventions,jfmk,jrfm,lubricants,neonatalscreening,neuroglia,particles,pharmaceutics,polymers,processes,technologies,viruses,vision

\section{Introduction}

The Fokker--Planck equation (FPE) describes the evolution of the probability density function (PDF) for
the state variables of dynamical systems modelled by stochastic differential equations (SDE).
Fokker--Planck equations are widely used for modelling stochastic phenomena in various fields, such as, for example, in physics, finance, biology, neuroscience, traffic flow \cite{schadschneider2010stochastic}. 
Yet, explicit closed-form solutions of FPE are rarely available~\cite{kumar2006solution}, especially in settings where the underlying dynamics is nonlinear. In particular, exact analytic solutions may be obtained only for a restricted class of systems following linear dynamics perturbed by white Gaussian noise, and for some nonlinear Hamiltonian systems~\cite{risken1996fokker, brics2013solve}.

Existing numerical approaches for computing Fokker--Planck solutions may be grouped into three broad categories: grid based, semi-analytical, and sample based methods.
The first category, comprises mainly finite difference and finite element methods~\cite{chang1970practical,pichler2013numerical}. These frameworks, based on   
integration of FPE employing numerical solvers for partial differential equations, entail computationally demanding calculations with inherent finite spatial resolution~\cite{leimkuhler2004simulating}.
%requiring fine discretisation steps to acquire numerically stable solutions. 
%In the meanwhile, they require substantial involvement from the researcher for constructing a stable numerical program suitable for the particular problem setting.

Conversely, semi-analytical approaches try to reduce the number of required computations by assuming conditional Gaussian structures~\cite{chen2018efficient}, or by employing cumulant neglect closures~\cite{linprobabilistic},  statistical linearisation \cite{roberts2003random,proppe2003equivalent}, or stochastic averaging \cite{grigoriu2013stochastic}. Although efficient for the settings they are devised for, their applicability is limited, since, the resulting solutions are imprecise or unstable in certain settings. 

 %Smoothed particles Hydrodynamics \cite{canor2013transient}
On the other hand, in the sample based category, Monte Carlo methods resort to stochastic integration of a large number of {\em independent} stochastic trajectories that as an ensemble represent the probability density \cite{oksendal2003,kroese2013handbook}. These methods are appropriate for computing {\em unbiased} estimates of exact expectations from empirical averages. % while positivity preserving advantage against classical methods
  Nevertheless, 
  %the required number of simulated stochastic trajectories scales exponentially for increasing dimension~\cite{robert2013monte,daum2003curse}, rendering the method unsuitable for high dimensional settings. Moreover,
  as we show in the following, cumulants of resulting distributions exhibit strong temporal fluctuations, when the number of simulated trajectories is not sufficiently large.

Surprisingly, there is an alternative sample based
approach built on {\em deterministic} particle dynamics. In this setting, the particles are not independent, but they rather {\em interact} via an (approximated)
probability density, and the FPE describes the mean field limit, when their number grows to infinity. 
This approach introduces a bias in the approximated expectations, but significantly reduces the variance for a given particle number.

Recent research, see e.g.~\cite{CCP19,sahani,reich2019fokker,liu2016kernelized},
has focused on particle methods
for models of thermal equilibrium, where the stationary density is known analytically. 
For these models, interacting particle methods have found interesting new applications in the field of probabilistic Bayesian inference: by treating the Bayesian posterior probability density as the stationary density of a FPE, the particle dynamics provides
posterior samples in the long time limit. For this approach, the particle dynamics is constructed
by exploiting the gradient structure of the probability flow of the FPE. This involves 
the relative entropy distance to the equilibrium density as a Lyapunov function.
Unfortunately, this structure does not apply to general
FPEs in \emph{non--equilibrium} settings, where the stationary density is usually unknown.

In this article, we introduce a framework for interacting particle systems that
may be applied to general types of Fokker--Planck equations.  Our approach is based on the fact that the instantaneous effective force on a particle due to diffusion is
proportional to the {\em gradient} of the {\em logarithm} of the exact probability {\em density} (GLD).
Rather than computing a differentiable estimate of this density (say by a kernel density estimator), we estimate the GLD directly without requiring knowledge of a stationary density. Thereby, we introduce an approximation to
the effective force acting on each particle, which becomes exact in the large particle number limit given the consistency of the estimator.

Our approach is motivated by recent developments in the field of 
machine learning, where GLD estimators have been studied independently and are
used
to fit probabilistic models to data. An application of these techniques to particle approximations 
for FPE is, to our knowledge, new.  \footnote{The approach in \cite{taghvaei2019accelerated} uses a GLD estimator 
different from ours for particle dynamics but with a probability flow towards equilibrium which 
is not given by a standard FPE.} Furthermore, our method provides also straightforward approximations of entropy production rates, which are of primary importance in 
non--equilibrium statistical physics~\cite{velasco2011entropy}.

This article is organised as follows: Section~\ref{sec:Deterministic_FP} describes the deterministic particle formulation
of the Fokker--Planck equation. Section~\ref{sec:gld} shows how a gradient of the logarithm of a density may be represented as the solution of a 
variational problem, while in Section~\ref{sec:estimators} we discuss an empirical approximation of the gradient-log-density. In Section~\ref{sec:Function}, we 
introduce function classes for which the variational problem may be solved explicitly, while in
Section~\ref{sec:expectations} we compare the temporal derivative of empirical expectations based on the 
particle dynamics with exact results derived from the Fokker--Planck equation.
Section~\ref{sec:equilibrium} is devoted to the class of equilibrium Fokker--Planck equations, where we discuss relations 
to Stein Variational Gradient Descent and other particle approximations of Fokker--Planck solutions. 
In Section~\ref{sec:statedep}, we show  
how our method may be extended
to general diffusion processes with state dependent diffusion, while Section~\ref{sec:Langevin} discusses how our framework may be employed to simulate second order Langevin dynamics.
In Section~\ref{sec:Results} we demonstrate various aspects of our method by simulating Fokker--Planck solutions for
different dynamical models. Finally, we conclude with a discussion and an outlook in Section~\ref{sec:outlook}.
\newpage

\section{Deterministic particle dynamics for Fokker--Planck equations} \label{sec:Deterministic_FP}
We consider Fokker--Planck equations of the type
\begin{equation}
\frac{\partial p_t(x)}{\partial t} =  -\nabla\cdot \left[f(x) p_t(x) - \frac{\sigma^2}{2} \nabla p_t(x))\right] 
\; .
\label{Fokker}
\end{equation}
Given an initial condition $p_0(x)$, Eq.~(\ref{Fokker})
describes the temporal development 
of the density $p_t(x)$ for the random variable $X(t) \in R^d$ 
following the stochastic differential equation
\begin{equation}
  dX(t)= f(X(t)) dt  + \sigma dB(t) \; .
  \label{eq:SDE}
\end{equation}
In Eq.~(\ref{eq:SDE}), $f(x)\in R^d$ denotes the drift function characterising the deterministic part of 
the driving force, while $dB(t)\in R^d$ represents the differential of a vector of independent Wiener processes capturing stochastic, Gaussian white noise excitations. For the moment, we restrict ourselves to state independent and diagonal {\em diffusion matrices}, i.e. 
 diffusion matrices  independent of  $X(t)$ (additive noise) with
diagonal elements $\sigma^2$ characterising the noise amplitude in each dimension. Extensions to more general settings are deferred to Section~\ref{sec:statedep}.

We may rewrite the FPE  Eq.~(\ref{Fokker}) in the form of a {\em Liouville} equation 
\begin{equation}
\frac{\partial p_t(x)}{\partial t} =  -\nabla\cdot \left[g(x,t)\; p_t(x) \right]  
\label{Liouville}
\end{equation}
for the {\em deterministic} dynamical system
\begin{equation}
\frac{dX}{dt} = g(X,t)\; ,  \qquad X(0) \sim p_0(x),
\label{particles1}
\end{equation}
(dropping the time argument in $X(t)$ for simplicity)
with velocity field
\begin{equation}
g(x,t) = f(x) - \frac{\sigma^2}{2} \nabla \ln p_t(x) \, .
\label{veloc1}
\end{equation}
Hence, by evolving an ensemble of $N$ {\em independent} realisations
of Eq.~(\ref{particles1}) (to be called 'particles' in the following) according to  
\begin{equation} 
\frac{dX_i}{dt} = g(X_i,t)\; , \qquad i=1,\ldots,N \qquad X_i(0) \sim p_0(x),
\label{particles2}
\end{equation}
we obtain an empirical approximation to the density $p_t(x)$.

Since the only source of randomness in Eq.~(\ref{particles1}) can be attributed to the initial conditions $X_i(0)$, averages computed from the particle approximation (Eq.~(\ref{particles2}))  are expected to have smaller variance compared to $N$ independent simulations of the SDE (Eq.~(\ref{eq:SDE})). 
Unfortunately, this approach requires perfect knowledge of the unknown instantaneous density $p_t(x)$ (c.f. Eq.~(\ref{veloc1})), that is actually the quantity we want to compute. 

Here, we circumvent this issue by introducing {\em statistical estimators} for the term $\nabla \ln p_t(x)$, computed from the entire ensemble $(X_1(t)),\ldots,X_N(t))$ of particles
at time $t$. Although this additional approximation introduces interactions among the particles via the estimator, for sufficiently large particle number $N$, fluctuations of the estimator are expected to be negligible and the limiting dynamics should 
converge to its mean field limit (Eq.~(\ref{particles1})) provided the estimator is asymptotically consistent. 
Thus, rather than computing a differentiable approximation to $p_t(x)$ from the particles, e.g. by a
kernel density estimator,  we show in the following section, how the function $\nabla \ln p_t(x)$ may be directly estimated from samples of $p_t(x)$.

%%%%%%%%%%%%%%%%%%%%%%%%%%%%%%%%%%%%%%

\section{Variational representation of gradient--log--densities} \label{sec:gld}

To construct a gradient--log--density (GLD) estimator we rely on a variational representation  introduced
by {\em Hyv\"arinen} in his {\em score--matching} approach for the estimation of non--normalised statistical models~\cite{hyvarinen2005estimation}. We favoured this approach over other
estimators \cite{li2017gradient,shi2018spectral}
 due to its flexibility to adapt to different function
classes chosen to approximate the GLD. 

Here, we use a slightly more general representation compared to \cite{hyvarinen2005estimation}
allowing for an extra arbitrary reference function  ${r(x) = (r^{(1)}(x),\ldots, r^{(d)}(x))}$ 
such that the component $\alpha$ of the gradient is represented as
\begin{equation}
\partial_\alpha \ln p(x) = r^{(\alpha)}(x) +
\arg\min_\phi\; {\cal{L}}^r_\alpha [\phi, p],
\label{esti1}
\end{equation}
where $\partial_\alpha \doteq 
\frac{\partial}{\partial x^{(\alpha)}}$ stands for the partial derivative with respect to
coordinate $\alpha$ of the vector ${x \equiv (x^{(1)}, \ldots  x^{(d)})}$.  

The cost function is defined as an expectation with respect to the density $p(x)$ by
\begin{equation}
{\cal{L}}^r_\alpha [\phi,p] = \int p(x) \left(\phi^2(x) + 2 r^{(\alpha)}(x) \phi(x) + 2 \partial_\alpha \phi(x)\right) dx \; ,
\label{costfun}
\end{equation}
with $dx$ representing the volume element in $R^d$. To obtain this relation, we use
integration by parts (assuming appropriate behaviour of densities and $\phi$ at
boundaries), and get 
\begin{eqnarray}
{\cal{L}}^r_\alpha [\phi,p]  & = &\int p(x) \left(\phi(x) + r^{(\alpha)}(x) - \partial_\alpha \ln p(x) \right)^2 dx  \nonumber \\
& -  & \int p(x) \left(\partial_\alpha \ln p(x)  - r^{(\alpha)}(x) \right)^2  dx.
\label{esti_min}
\end{eqnarray}
Minimisation with respect to $\phi$ yields Eq.~(\ref{esti1}).  
%%%%%%%%%%%%%%%%%%%%%%%%%%%%%%%%%%%%%%%%%%%%%%%%%%%%%%%%%%%%%%%%%%%
\section{Gradient--log--density Estimator} \label{sec:estimators}
To transform the variational formulation into a GLD estimator based on $N$ sample points $(X_1,\ldots, X_N)$, we replace the density $p(x)$ in Eq.~(\ref{costfun}) by the empirical 
distribution $
{\hat{p}_t(x) = \frac{1}{N}\sum_{i=1}^N \delta (x - X_i(t))}
$, i.e.
\begin{equation}
{\cal{L}}^r_\alpha[\phi,p_t] \approx {\cal{L}}^r_\alpha[\phi, \hat{p}_t] =  \frac{1}{N} \sum_{i=1}^N \left(\phi^2(X_i) + 2 r^{(\alpha)}(X_i) \phi(X_i)+ 2 \partial_\alpha \phi(X_i)\right)\, , 
\label{cost_empir} 
\end{equation}
and
\begin{equation}
\partial_\alpha \ln p_t(x)  \approx r^{(\alpha)}  +
\arg\min_{\phi\in {\cal{F}}}\; {\cal{L}}^r_\alpha[\phi, \hat{p}_t]\; ,
\label{esti_general}
\end{equation}
where ${\cal{F}}$ is an appropriately chosen family of functions with controllable complexity.
By introducing the estimator of Eq.~(\ref{esti_general}) in Eq.~(\ref{particles2}), we obtain a particle representation for the Fokker--Planck equation
\begin{equation}
\frac{dX_i^{(\alpha)}}{dt} = f^{(\alpha)}(X_i) - \frac{\sigma^2}{2}\left (r^{(\alpha)}(X_i) + \arg\min_{\phi\in {\cal{F}}}\; {\cal{L}}^r_\alpha[\phi, \hat{p}_t]\right),
\label{dynamics_general} 
\end{equation}
for $i=1,\ldots,N\; \mbox{and} \; \alpha =1,\ldots,d $, with
 
\begin{equation} 
\hat{p}_t(x) =   \frac{1}{N}\sum_{i=1}^N \delta (x - X_i)\, ,\qquad  X_i(0) \sim p_0(x).
\nonumber
\end{equation}
Although, in this article, we use $r\equiv 0$ for all simulated examples, the choice $r(x) = \frac{2}{\sigma^2} f(x)$,
which cancels the first two terms in Eq.~(\ref{dynamics_general}),
leads to interesting relations with other particle approaches for simulating Fokker--Planck solutions for equilibrium systems (c.f. Section~\ref{sec:equilibrium}).% [DISCUSS IN CONCLUSSION]

\subsection{Estimating the entropy rate}
Interestingly, the variational approach provides us with a simple, built in method for computing the 
entropy rate (temporal change of entropy) of the stochastic process 
(Eq.~(\ref{eq:SDE})).

Using the FPE (\ref{Fokker})
and integration by parts one can derive the well known relation, see e.g.~\cite{tome2015stochastic},
\begin{equation}
- \frac{d}{dt} \int p_t(x) \ln p_t(x) dx =  \frac{\sigma^2}{2} \sum_{\alpha =1}^d
\int p_t(x) \left(\partial_\alpha \ln p_t(x)  \right)^2 dx + \int p_t(x) \nabla\cdot f(x) dx\,.
\label{entro_rate}
\end{equation}
The first term on the right hand side is usually called entropy production, whereas the second term
corresponds to the entropy flux. In the stationary state, the total entropy rate vanishes.
For equilibrium dynamics, both terms vanish individually at stationarity.
This should be compared to the minimum of the cost function (Eq.~(\ref{esti_min})) which
for $r\equiv 0$ equals
\begin{equation}
\min_{\phi} {\cal{L}}^0_\alpha [\phi,p_t] =  -  \int p_t(x) \left(\partial_\alpha \ln p_t(x)  \right)^2  dx\,.
\end{equation}
Thus we obtain the estimator
\begin{equation}
- \frac{d}{dt} \int p_t(x) \ln p_t(x) dx \approx  - \frac{\sigma^2}{2} \sum_{\alpha =1}^d
\min_{\phi} {\cal{L}}^0_\alpha [\phi,\hat{p}_t] + \frac{1}{N}
\sum_{i=1}^N \nabla\cdot f(X_i) \, .
\end{equation}
We will later see for the case of equilibrium dynamics that a similar method may be employed to
approximate the relative entropy distance to the equilibrium density.
%%%%%%%%%%%%%%%%%%%%%%%%%%%%%%%%%%%%%%%%%%%%%%%%%%%%%%%%%%%%%%%
\section{Function classes} \label{sec:Function}
In the following, we discuss choices for families of functions ${\cal{F}}$ leading to explicit, closed
form solutions for estimators. 
\subsection{Linear models}
A simple possibility is to choose linearly parametrised functions of the form
\begin{equation}
\phi(x) = \sum_{l=1}^m a_k \phi_k(x)\, ,
\label{linpar}
\end{equation}
where the $\phi_k(x)$ are appropriate basis functions, e.g.~polynomials, radial basis
functions or trigonometric functions. For this linear parametrisation,  the empirical cost (Eq.~(\ref{cost_empir})) is quadratic in the parameters $a_k$ and can be minimised explicitly.
A straightforward computation shows that 
\begin{eqnarray}
\frac{dX_i}{dt} = f(X_i) - \frac{\sigma^2}{2} r(X_i) +  
\frac{\sigma^2}{2} \sum_{k,j=1}^m (C^{-1})_{k j} \phi_k(X_i)\sum_{l=1}^N 
\left\{\nabla \phi_j(X_l) + \phi_j(X_l) r(X_l) \right\} \, ,
\label{dynamics_basis}
\end{eqnarray} 
with $ \displaystyle C_{kl} = \sum_{i=1}^N \phi_k(X_i) \phi_l(X_i)$.

Obviously, we require the number of samples to be greater than the number of employed basis functions, i.e. $N\geq m+1$, to have a non--singular matrix $C$. This restriction 
can be lifted by introducing an additional penalty for regularisation. Eq.~(\ref{dynamics_basis})
is independent of the reference function $r$, when $r$ belongs to the linear span of the selected basis functions.
However, this model class with a finite parameter number has limited complexity. Thus, 
even when the sample number $N$ grows large, we do not expect, in general, convergence to the 
mean field limit.

\subsection{Kernel approaches}
Here, we consider a family ${\cal{F}}$ of functions for which the effective number
of parameters to be computed is not fixed beforehand, but rather increases with the sample number $N$: a  {\em reproducing kernel Hilbert space} (RKHS) of functions defined by
a positive definite (Mercer) kernel $K(\cdot,\cdot)$. Statistical models based on such function spaces have played a prominent role in the field of machine learning in recent years~\cite{shawe2004kernel}.  

A common, kernel based approach to regularise the minimisation
of empirical cost functions is via penalisation using the RKHS norm $\| \cdot\|_{\mbox{\tiny{RKHS}}}$
of functions in ${\cal{F}}$. This can also be understood as penalised version of 
a linear model (\ref{linpar}) with infinitely many feature functions $\phi_k$.
For so called universal kernels~\cite{scholkopf2001learning} this unbounded complexity 
suggests that we could expect asymptotic convergence of the GLD estimator
(see~\cite{sutherland2017efficient} for related results) and a corresponding convergence of the particle model to its mean field limit. However, a rigorous proof may not be trivial, since particles in our setting are not independent. 

The explicit form of the kernel based approximation is given by
\begin{equation}
\partial_\alpha \ln p(x) \approx  r^{(\alpha)} (x) +
\arg\min_{\phi\in {\cal{F}}}\; \left\{{\cal{L}}^r_\alpha[\phi, \hat{p}] +  \frac{\lambda}{N} \| \phi \|_{\mbox{\tiny{RKHS}}}^2\right\}\, ,
\label{penal_cost}
\end{equation}
where the parameter $\lambda$ controls the strength of the penalisation. Again, this 
optimisation problem can be solved
in closed form in terms of matrix inverses. One can prove  
a {\em representer theorem} which states that the minimiser $\phi(x)$ in
Eq.~(\ref{penal_cost}) is a  linear combination of kernel functions evaluated at the 
sample points $X_i$, i.e.,
\begin{equation}
\phi(x) = \sum_{i=1}^N a_i K(x,X_i) \, .
\label{kernel_rep}
\end{equation} 
For such functions, the RKHS norm is given by
\begin{equation}
\| \phi \|_{\mbox{\tiny{RKHS}}}^2 = \sum_{i,j=1}^N a_i a_j K(X_i,X_j) \, .
\label{RKHSnorm}
\end{equation}
Hence, this representation leads again to a quadratic form in the $N$ coefficients.

A short computation yields
\begin{equation}
a_j = - \sum_{k=1}^N \left((K^2 + \lambda K)^{-1}\right)_{jk} \sum_{l=1}^N \left\{\partial_{\alpha_l} K(X_l, X_k) 
+ K(X_l, X_k)  r^{(\alpha)} (X_l)\right\}\, ,
\label{kernel_coeff}
\end{equation}
where $K_{ij} \doteq K(X_i,X_j)$.
Similar approaches for kernel based GLD estimators have been discussed 
in \cite{li2017gradient,shi2018spectral}. For $r=0$, Eq.~(\ref{kernel_coeff}) agrees with the GLD estimator of~\cite{li2017gradient} derived
by inverting {\em Stein's} equation, or by minimising the {\em Kernelised Stein discrepancy}.

The resulting particle dynamics is given by
\begin{equation}
\frac{dX_i}{dt} = f(X_i) - \frac{\sigma^2}{2} r(X_i) +  
\frac{\sigma^2}{2} \sum_{k=1}^N \left((K + \lambda I)^{-1}\right)_{i k}\sum_{l=1}^N 
\left\{\nabla_l K(X_l, X_k) + K(X_l, X_k) r(X_l) \right\}\, .
\label{dynamics_kernel}
\end{equation} 
Note that here also the inverse matrix depends on the 
particles $X_k$. 
In the limit of small $\lambda$, the right hand side becomes independent
of the reference function $r$. 

In the present article, we employ Gaussian radial basis function 
(RBF)  kernels given by
\begin{equation}
K(x,x') = \exp\left[-\frac{1}{2\,l^2}\| x- x'\|^2\right]\, ,
\end{equation} 
with a length scale $l$.    A different possibility would be given
by kernels with a {\em finite dimensional} feature representation 
\begin{equation}
K(x,x') = \sum_{j=1}^m \phi_j(x)  \phi_j(x') \, ,
\end{equation} 
which may also be interpreted as a {\em linear model} as in Eq.~\ref{linpar} with a $L_2$ penalty
on the unknown coefficients.
%%%%%%%%%%%%%%%%%%%%%%%%%%%%%%%%%%%%%%%%%%%%%%%%%%%%%%%%%
\subsection{A sparse kernel approximation}
The inversions of the $N\times N$ matrices in Eq.~(\ref{dynamics_kernel}) have to be performed at each step of a time discretised ODE system (Eq.~(\ref{dynamics_kernel})).
For large $N$, the cubic complexity could become too time consuming. Hence, here, we resort to a  
well established approximation in machine learning to overcome this issue, by
applying a sparse approximation to the optimisation problem of Eq.~(\ref{penal_cost}), see e.g.~\cite{rasmussen2003gaussian}. In particular, we introduce a smaller set of $M\ll N$ {\em inducing points} $\{z_k\}_{k=1}^M$, that need not necessarily be a subset of the $N$ particles. We then minimise the penalised cost function (Eq.~\ref{penal_cost}) in the finite dimensional family of functions 
\begin{equation}
\phi(x) = \sum_{i=1}^M a_i  K(x,z_i)  \, .
\label{kernel_rep_sparse}
\end{equation} 
This may also be understood as a special linear parametric approximation.
To keep matrices well conditioned, in practice we add a small 'jitter' 
term to Eq.~(\ref{penal_cost}), i.e., we use 
\begin{equation}
\lambda \| \phi \|_{\mbox{\tiny{RKHS}}}^2 + \epsilon \| \phi \|_2^2 \, ,
\end{equation} 
as the total penalty. 
In the limit $\lambda, \epsilon\to 0$, this representation reduces to an approximation 
of the form of Eq.~(\ref{linpar})
with $M$ basis functions $K(\cdot, z_l)$ for $l=1,\ldots,M$. 

By introducing the matrices 
\begin{equation}
K^{zz}_{kl} \doteq K(z_k, z_l) + \epsilon \delta_{kl} \qquad
K^{xz}_{ij} \doteq K(X_i, z_j) \, ,
\end{equation}
and
\begin{equation} \label{eq:A}
A \doteq K^{xz} \left[ (\lambda + \epsilon) I + 
(K^{zz})^{-1} (K^{xz})^\top (K^{xz})\right]^{-1}(K^{zz})^{-1}\, ,
\end{equation} 
%{\bf Better formulation:} Consider $\epsilon = 0$ from the start and use instead ?????
%\begin{equation}
%A \doteq K^{xz} \left[ \lambda K^{zz} +  (K^{xz})^\top (K^{xz})\right]^{\dagger}
%\end{equation} 
we replace the particle dynamics of Eq.~(\ref{dynamics_kernel}) by
\begin{equation}
\frac{dX_i}{dt} = f(X_i) - \frac{\sigma^2}{2} r(X_i) +  
\frac{\sigma^2}{2} \sum_k A_{i k}\sum_l 
\left\{\nabla_l K(X_l, z_k) + K(X_l, z_k) r(X_l) \right\} .
\label{dynamics_kernel_sparse}
\end{equation} 
%where $B^{\dagger}$ denotes the Moore--Penrose pseudo--inverse of matrix $B$.
Hence, for this  approximation we have to invert only $M\times M$ matrices.
For fixed $M$, the complexity of the GLD estimator is limited.
Results for log--density--estimators in machine learning (obtained for independent data) indicate that for a moderate growth of the number of inducing points 
$M$ with the number of particles $N$, similar approximation rates may be obtained
as for full kernel approaches.
%%%%%%%%%%%%%%%%%%%%%%%%%%%%%%%%%%%%%%%%%%%%%%%%%%%%%%%%%%%%%%%%%%%%%%%%%%%
\section{A note on expectations} \label{sec:expectations}
In this section we present a preliminary discussion of the quality of the particle method
to approximate expectations of scalar functions $h$ of the random variable $X(t)$. We 
concentrate on the temporal development of $h(X(t))$. 
While it would be important to obtain an estimate of the approximation error
over time, we will defer such an analysis to future publications and only concentrate 
on a result for the first time derivative of expectations, i.e. the evolution 
over infinitesimal times.

Using the FPE (Eq.~(\ref{Fokker}))
and integrations by part one derives the exact result
\begin{equation}
\frac{d \langle h(X) \rangle}{dt} = \langle L_x h(X)\rangle \, ,
\label{expec_evolution_exact}
\end{equation}
where $\langle \cdot \rangle $ denotes the expectation with respect to $p_t(x)$
and the operator $L_x$ equals the {\em generator} of the process, i.e.,
\begin{equation}
L_x \doteq f(X)\cdot \nabla + \frac{\sigma^2}{2}\nabla^2 \, .
\label{generator}
\end{equation}
To obtain 
a related result for empirical expectations based on particles, we  
employ the relation
\begin{equation}
\frac{d h(X_i)}{dt}  = \nabla h(X_i)\cdot \frac{d X_i}{dt}\, 
\label{expec_deriv}
\end{equation}
and a direct computation using the dynamics of Eq.~(\ref{dynamics_basis}) 
and Eq.~(\ref{dynamics_kernel})
yields the result
\begin{equation}
\frac{d \langle h(X) \rangle_{\hat{p_t}}}{dt} = \langle L_x h(X)\rangle_{\hat{p_t}} + \Delta \, ,
\label{expec_evolution_part}
\end{equation}
where $\langle\cdot \rangle_{\hat{p_t}}$ denotes expectation with respect to the empirical distribution $\hat{p_t}$
of the particles. Hence, if the remainder $\Delta$ is small, the change of empirical particle 
averages should not deviate much from the corresponding exact ones.
This remainder term is given by
\begin{equation}
\Delta = \frac{\sigma^2}{2} \langle\left(r(X) + \nabla\right)\cdot \left(\hat{\nabla} h(x) - \nabla h(x)\right)\rangle_{\hat{p_t}} \, ,
\end{equation}
where $\hat{\nabla}h(x)$ stands for the approximation of the vectorial function $\nabla h(x)$ 
based on the 'data' $\nabla h(X_l)$ using regression with a linear combination of basis functions $\phi_h(x)$ or by regularised kernel  regression. The explicit formulas for the two cases are 
\begin{equation}
\hat{\nabla} h(x) = \sum_{j,k=1}^M \phi_j(x)  \left(C^{-1}\right)_{j k} \sum_l \phi_k(x_l) \nabla h(X_l)
\end{equation}
and
\begin{equation}
\hat{\nabla} h(x) = \sum_{j,k=1}^N K(x,X_j) \left((K + \lambda I)^{-1}\right)_{j k} \nabla h(X_k) \, ,
\end{equation}
respectively. If $\nabla h(x)$ is well 
approximated by basis functions, the remainder $\Delta$ is small. 
 If indeed ${\nabla h(x) = \sum_{n=1}^M c_n \phi_n(x)}$,  for some $c_n\in R^d$,
the remainder term vanishes, $\Delta =0$.  By its similarity to the finite basis function model, this result should also be valid for the sparse kernel dynamics of Eq.~(\ref{dynamics_kernel_sparse}), when the penalty $\lambda$ is small. 
One might conjecture that the temporal development of expectations for reasonably smooth functions might be faithfully represented by the particle dynamics. 
This conjecture is supported by our numerical results.
%%%%%%%%%%%%%%%%%%%%%%%%%%%%%%%%%%%%%%%%%%%%%%%%%%%%%%%%%%%%%%%%%%%%%%%%%%%%%
\section{Equilibrium dynamics}  \label{sec:equilibrium}
An important class of stochastic dynamical systems describe {\em thermal equilibrium}, for
which the drift function $f$ is the negative gradient of a potential $U$, while the limiting equilibrium density $p_\infty$ is explicitly given by a Gibbs distribution:
\begin{eqnarray}
f(x) = - \nabla U(x) \\
\nabla \ln p_\infty (x) = \frac{2}{\sigma^2} f(x)  .
\end{eqnarray}
For this class of models, our method provides a simple and built in estimator for the relative entropy
between the instantaneous, $p_t$, and the equilibrium density, $p_\infty$. As we discuss here, our framework may also be related to two other particle 
approaches, that converge to the (approximate) equilibrium density. 
\subsection{Relative entropy}
The relative entropy or {\em Kullback--Leibler divergence}
 is defined as
\begin{equation}
D(p_t | p_\infty) \doteq \int p_t(x) \ln\frac{p_t(x)}{p_{\infty}(x)} dx\, .
\end{equation}
Following a similar calculation that led to Eq.~(\ref{entro_rate}), we obtain
\begin{eqnarray}
\frac{d}{dt}  D(p_t | p_{\infty}) = - \frac{\sigma^2}{2} \int p_t(x) \left\|\nabla \ln p_t(x) - \nabla \ln p_\infty(x)
\right\|^2 dx 
\nonumber \\
= - \frac{2}{\sigma^2} \int p_t(x) \| g(x,t) \|^2 dx \, ,
\label{KL_deriv}
\end{eqnarray}
where $g(x,t)$ indicates the velocity field of the particle system defined in Eq.~(\ref{particles1}).
The first equality holds for arbitrary drift functions. To obtain the second equality, we have 
inserted the explicit result for $p_\infty$.  

Hence, we may compute the relative entropy at any time $T$ as a time integral
\begin{equation}
D(p_T | p_{\infty}) = D(p_0 | p_{\infty}) -\frac{2}{\sigma^2} \int_0^T  \left\{\int p_t(x) \| g(x,t) \|^2 dx
\right\} dt\, ,
\end{equation}
where the inner expectation is easily approximated by our particle algorithm. This result shows that the exact velocity field $g(x,t)$ converges to 0 for $t \to\infty$ and one expects 
particles to also converge to fixed points. For other, non--equilibrium systems asymptotic fixed points are, however, the exception.

%%%%%%%%%%%%%%%%%%%%%%%%%%%%%%%%%%%%%%%%%%%%%%%%%%%%%%%%%%%%%%%%%%%%%
\subsection{Relation to Stein Variational Gradient Descent}
Recently, {\em Stein variational gradient descent} (SVGD), 
a kernel based particle algorithm, has attracted considerable attention 
in the machine learning community \cite{liu2016stein,liu2017stein}. The algorithm is designed to provide approximate
samples from a given density $p_{\infty}$ as the asymptotic fixed points of a deterministic 
particle system. Setting $- \ln p_{\infty}(x) = U(x) + \mbox{const}$, 
SVGD is based on the dynamics 
\begin{equation}
\frac{dX_i}{dt} = \sum_l \left\{ - K(X_i, X_l) \nabla U(X_l)  + \nabla_l K(X_i, X_l) \right\} \; .
\label{dynamics_Stein}
\end{equation}
This can be compared to our approximate FPE dynamics 
(Eq.~(\ref{dynamics_kernel})) for the equilibrium case by setting $\sigma^2 =2$
and $r(x) = f(x) = -\nabla U(x)$. For this setting, both algorithms have in fact, the
same conditions   
\begin{equation}
\sum_l \left\{ - K(X_i, X_l) \nabla U(X_l)  + \nabla_l K(X_i, X_l) \right\} = 0\, ,
\end{equation}
for the 'equilibrium' fixed points. See~\cite{liu2016kernelized} for a discussion of these fixed points 
for different kernel functions. 
However, both dynamics differ for finite times $t$, 
where a single time step of SVGD is computationally simpler, being free of the matrix inversion required by our framework. 
The mean field limit $N\to\infty$ of Eq.~(\ref{dynamics_Stein}) differs from the FPE, and the resulting  partial differential equation is nonlinear~\cite{nusken2019note}.
Nevertheless, it is possible to interpolate between the two particle dynamics. In fact, in the limit of a large regularisation parameter $\lambda\to\infty$, the inverse matrix
in Eq.~(\ref{dynamics_kernel}) becomes diagonal, i.e.
$(K + \lambda I)^{-1}\simeq \frac{1}{\lambda} I$, and we recover SVGD
(Eq.~(\ref{dynamics_Stein})) by introducing a rescaled time $\tau \doteq  t/\lambda $.
This result could be of practical importance when the goal is to approximate the 
stationary distribution, irrespectively of the finite time dynamics. 
The SVGD combines faster matrix operations with slower relaxation times to equilibrium 
compared to the FPE dynamics.  It would be interesting to see, if an optimal computational 
speed of a particle algorithm might be achieved at some intermediate 
regularisation parameter $\lambda$.
%%%%%%%%%%%%%%%%%%%%%%%%%%%%%%%%%%%%%%%%%%%%%%%%%%%%%%%%%%%%%%%%%%%%%%%%%%
\subsection{Relation to geometric formulation of FPE flow}
%%%%%%%%%%%%%%%%%%%%%%%%%%%%%%%%%%%%%%%%%%%%%%%%%%%%%%%%%%%%%%%%%%%%%%%%%%%%
Following Otto \cite{otto2001geometry} and Villani \cite{villani2008optimal}, the FPE for the equilibrium case can be viewed as a gradient flow on the manifold of probability densities with respect to the Wasserstein metric. This formulation can be used to define an implicit Euler
time discretisation method for the dynamics of the density $p_t$. For small times $\delta t$ (and $\sigma^2 = 2$) 
this is given by the variational problem
\begin{equation}
p_{t+\delta t} = \arg\inf_p \left( W_2^2(p, p_t) +  \delta t D(p \| p_\infty) \right)
\label{Wasser_flow}
\end{equation}
in terms of the Kullback--Leibler divergence and the
$L_2$ {\em Wasserstein distance} $\mathcal{W}_2$. The latter gives the minimum 
of $\langle \|X - X(t)\|^2 \rangle$ for two random variables $X(t)$ and $X$ where the expectation 
is over the joint distribution with fixed marginals $p_t$ and $p$.
Using the dual formulation for a regularised Wasserstein distance, approximate numerical algorithms for solving Eq.~(\ref{Wasser_flow}) have been developed by \cite{frogner2018approximate}
and by~\cite{caluya2019gradient} with applications to simulations of FPE.

We show in the following that  Eq.~(\ref{Wasser_flow}) may be cast into a form 
closely related to our variational formulation (Eq.~(\ref{esti1})) for $r(x) = f(x)$. 
Assuming that $X$ and $X(t)$ are related
through a deterministic (transport) mapping of the form
\begin{equation}
X = X(t) + \delta t \nabla \psi(X(t)) \;, 
\end{equation}
we may represent the Wasserstein distance in terms of $\psi$ 
and the variational problem may be rewritten as
\begin{equation}\label{var_wasser}
p_{t+ \delta t}(x) = p_t(x) - \delta t \nabla \left(p_t(x) \nabla \psi^*(x)\right),
\end{equation}
where
\begin{equation}
\psi^*= \arg\min_{\nabla \psi} \frac{\delta t^2}{2} \int \|\nabla \psi(x) \|^2 p_t(x) dx + \delta t D(p_{t+dt} \| p_\infty)\, .
\end{equation} 
To proceed, we 
expand the relative entropy to first order in $\delta t$, inserting the representation Eq.~(\ref{var_wasser}) for $p_{t+ \delta t}(x)$, obtaining thereby 
\begin{eqnarray}
\frac{\delta t}{2} \int \|\nabla \psi(x) \|^2 p_t(x) dx + D(p_{t+\delta t} \| p_\infty) =
D(p_{t} \| p_\infty) +\nonumber  \\
 + \frac{\delta t}{2} 
\left(\int p_t(x) \left\{ \|\nabla \psi(x) \|^2 - 2\nabla^2 \psi(x) + 2 \nabla U(x)\cdot \nabla \psi(x) \right\} dx \right) + \\ + O(\delta t^2).
\label{Wasser_expand}
\nonumber
\end{eqnarray} 
Minimisation ignoring the $O(\delta t^2)$ terms (employing integration by parts) yields
\begin{equation}
\nabla \psi^*(x) = -\nabla U(x) -  \nabla \ln p_t(x),
\end{equation}
which is closely related to our cost function Eq.~(\ref{costfun}), if we identify $\phi(x) = - \nabla \psi(x)$. By replacing $p_t$ by samples, 
the empirical cost function may be regularised with a RKHS norm penalty resulting in a nonparametric estimator for unnormalised log--density 
$\psi^*(x) = - \ln p_t(x) - U(x) +\mbox{const}$ as shown in \cite{batz2016variational}. 
One could use this estimator as an alternative to our approach. This would lead 
to a simultaneous estimate of all components of the GLD. In our approach, each of the $d$ components of the gradient is computed individually. In this way, we avoid additional second derivatives of kernels, which would increase the dimensionality of the resulting matrices. 
%%%%%%%%%%%%%%%%%%%%%%%%%%%%%%%%%%%%%%%%%%%%%%%%%%%%%%%%%%%%%%%%%%%%%%%%%%%%%%%%%%%%
\section{Extension to general diffusion processes} \label{sec:statedep}
The Fokker--Planck equations for an SDE with arbitrary drift $f(x)$ and general, state dependent 
diffusion matrix $D(x)$ is given by
\begin{equation}
\frac{\partial p_t(x)}{\partial t}  =  \nabla\cdot \left[- f(x) p_t(x) + \frac{1}{2} \nabla\cdot (D(x)\; p_t(x))\right]  .
\label{FPgeneral}
\end{equation}
This may again be written in the form of a Liouville equation (Eq.~(\ref{Liouville})) where the effective force term equals 
\begin{equation}\label{eq:statedep}
g(x,t) = f(x) - \frac{1}{2} \nabla\cdot D(x)  - \frac{1}{2} D(x) \nabla \ln p_t(x).
\end{equation}
%We will not study the most general case here, but concentrate in the next section on a 2nd order Langevin dynamics (aka known as Karmer's equation).

%%%%%%%%%%%%%%%%%%%%%%%%%%%%%%%%%%%%%%%%%%%%%%%%%%%%%%%%%%%%%%%%%%%%%%%%%%%%%%%%%%%%
\section{Second order Langevin dynamics (Kramer's equation)}\label{sec:Langevin}
For second order Langevin equations, the system state comprises positions $X\in R^d$ and velocities $V\in R^d$ following the coupled SDE 
\begin{align}\label{eq:Lang}
dX &= V dt \\ 
dV &= \left(- \gamma V + f(X)\right) dt + \sigma dB_t\;.
\end{align} 
In Eq.~(\ref{eq:Lang}), the dynamics describe the effect of a friction force, $ \gamma V $, an external force, $f(X)$, and a fluctuating force, where $\gamma$ denotes the dissipation constant. 
In this setting, the effective
{\em deterministic} ODE system is given by
\begin{align} 
\frac{dX}{dt}& = V  \nonumber \\
\frac{dV }{dt}&= - \gamma V + f(X)  - \frac{\sigma^2}{2} \nabla_v \ln p_t(X,V) \;.
\label{Kramers_det}
\end{align} 
Considering here the equilibrium case, we set $f(x) = - \nabla U(x)$ for which the stationary density
equals 
\begin{equation}
\ln p_\infty(X,V) = -  \beta\left(\frac{\|V\|^2}{2} + U(X)\right) \equiv - \beta H(X,V),
\end{equation}
where $\beta = \frac{2\gamma}{\sigma^2}$ and
$H(x,v) = \frac{\|V\|^2}{2} + U(x)$ denotes the {\em Hamiltonian} function. 
Inserting $p_\infty$ into Eq.~(\ref{Kramers_det}), we find that 
for $t\to\infty$, the damping and the 
density dependent part of the force cancel and we are left with pure Hamiltonian dynamics
\begin{align}
\frac{dX}{dt} &= V  \nonumber \\
\frac{dV}{dt}&=  - \nabla U(X),
\end{align} 
for which all particles become completely {\em decoupled}, with each one conserving energy
separately. Of course, this result also precludes fixed point solutions to the particle dynamics. 

The asymptotic behaviour is also reflected in the expression for 
the change of the relative entropy for Kramer's equation. Similar to Eq.~(\ref{KL_deriv}) we obtain
\begin{eqnarray}
\frac{d}{dt}  D(p_t | p_{\infty}) = - \frac{\sigma^2}{2} \int p_t(x,v) \left\|\nabla_v \ln p_t(x,v) - \nabla_v \ln p_\infty(x,v) 
\right\|^2 dx dv \nonumber \\
= - \frac{2}{\sigma^2} \int p_t(x,v) \| \gamma v + \frac{\sigma^2}{2} \nabla_v \ln p(x,v) \|^2 dx dv.
\label{KL_deriv_Kram}
\end{eqnarray}
When the system approaches equilibrium, both terms in the norm cancel out and 
the entropy production rate converges to $0$.

\newpage

\section{Simulating accurate Fokker--Planck solutions for model systems} \label{sec:Results}

To demonstrate the accuracy of our approach, we simulated solutions of FPEs for a range of model systems and compared the results with those obtained from direct stochastic simulations (Monte Carlo sampling) of same particle number, and analytic solutions, where relevant. We tested our framework on systems with diverse degrees of nonlinearity and dimensionality, as well as with various types of noise (additive/multiplicative). We quantified the accuracy of transient and steady state solutions resulting from our method in terms of 1-Wasserstein distance~\cite{villani2008optimal} and Kullback Leibler (KL) divergence (Appendix~\ref{sec:KL} and~\ref{sec:Wasser}), along with squared error of distances between distribution cumulants. 
For evaluating particle solutions for nonlinear processes, where analytical solutions of the Fokker--Planck equation are intractable, we simulated a very large number ($N^{\infty}$) of stochastic trajectories that we considered as ground truth Fokker--Planck solutions. We employed an Euler--Maruyama and forward Euler integration scheme of constant step size $dt=10^{-3}$ for stochastic and deterministic simulations respectively.  

\subsection{Linear conservative system with additive noise} \label{sec:2DOU}

\begin{figure}[htbp]
  \centering
  \includegraphics[width=0.95\textwidth]{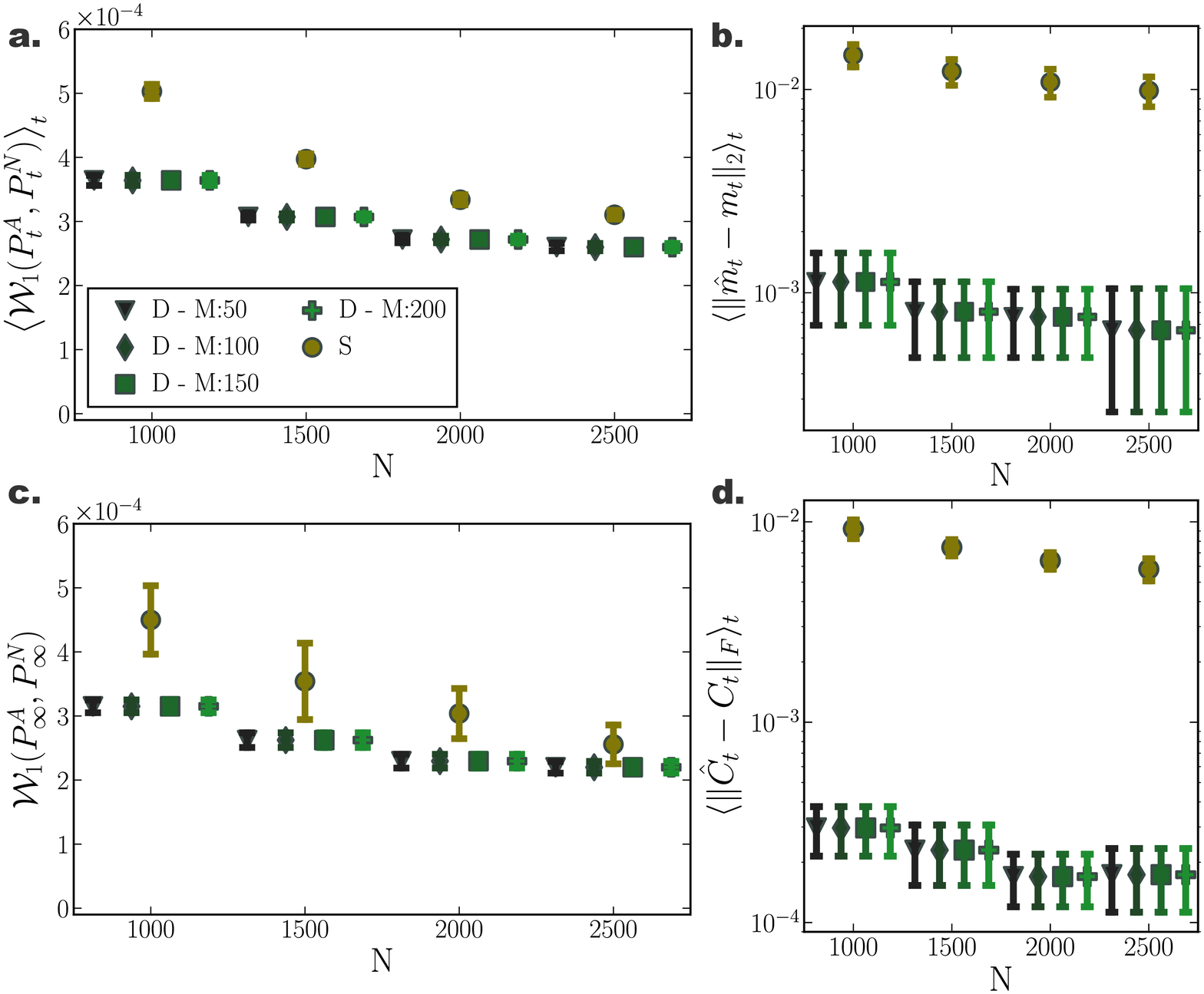}
  \caption{\textbf{Accuracy of Fokker--Planck solutions for two dimensional Ornstein Uhlenbeck process. }   \textbf{(a.)} Mean, $\left< \mathcal{W}_1(P_t^{A},P_t^N) \right>_t$, and \textbf{(c.)} stationary $ \mathcal{W}_1 (P_{\infty}^{A},P_{\infty}^N)$,  1-Wasserstein  distance,  between analytic solution and deterministic(D)/stochastic(S) simulations of $N$ particles (for different inducing point number $M$).  \textbf{(b.)} Average temporal deviations from analytic mean $m_t$ and \textbf{(d.)} covariance matrix $C_t$ for deterministic and stochastic system for increasing particle number $N$. Deterministic particle simulations consistently outperformed stochastic ones in approximating the temporal evolution of the mean and covariance of the distribution for all examined particle number settings.
  (Further parameter values: regularisation constant $\lambda=0.001$, Euler integration time step $dt=10^{-3}$, and RBF kernel length scale $l$ estimated at every time point as two times the standard deviation of the state vector. Inducing point locations were selected randomly at each time step from a uniform distribution spanning the state space volume covered by the state vector.)}
  \label{fig:2d_OUb}
\end{figure}

For a two dimensional Ornstein-Uhlenbeck process (Appendix~\ref{ap:OU2}) transient and stationary densities evolved through deterministic particle simulations (D) consistently outperformed their stochastic counterparts (S) comprising same number of particles in terms accuracy in approximating the underlying density (Fig.~\ref{fig:2d_OUb}). In particular, comparing the 1-Wasserstein distance between samples from analytically derived densities ($P_t^A$) (Appendix~\ref{ap:SolOU}) - considered here to reflect the ground truth - and the deterministically (D) or stochastically (S) evolved densities ($P_t^N$),
$\mathcal{W}_1(P_t^{A},P_t^N)$, we observed smaller Wasserstein distances to ground truth for densities evolved according to our deterministic particle dynamics, both for transient (Fig.~\ref{fig:2d_OUb}(a.)) and stationary (Fig.~\ref{fig:2d_OUb}(c.)) solutions. Specifically, we quantified the transient deviation of simulated densities from ground truth by the average temporal 1-Wasserstein distance, $\left< \mathcal{W}_1(P_t^{A},P_t^N) \right>_t$. For small particle number, deterministically evolved interacting particle trajectories represented more reliably the evolution of the true probability density compared to independent stochastic ones, as portrayed by smaller average Wasserstein distances. For increasing particle number the accuracy of the simulated solutions with the two approaches converged. Yet, although for $N=2500$ particles the stochastically evolved densities suggest \emph{on average} (over trials) comparable approximation precision with their deterministic counterparts, the deterministically evolved densities delivered more reliably densities of a certain accuracy, as proclaimed by the smaller dispersion of Wasserstein distances among different realisations (Fig.~\ref{fig:2d_OUb}(a., c.)).

\begin{figure}[htbp]
  \centering
  \includegraphics[width=0.7\textwidth]{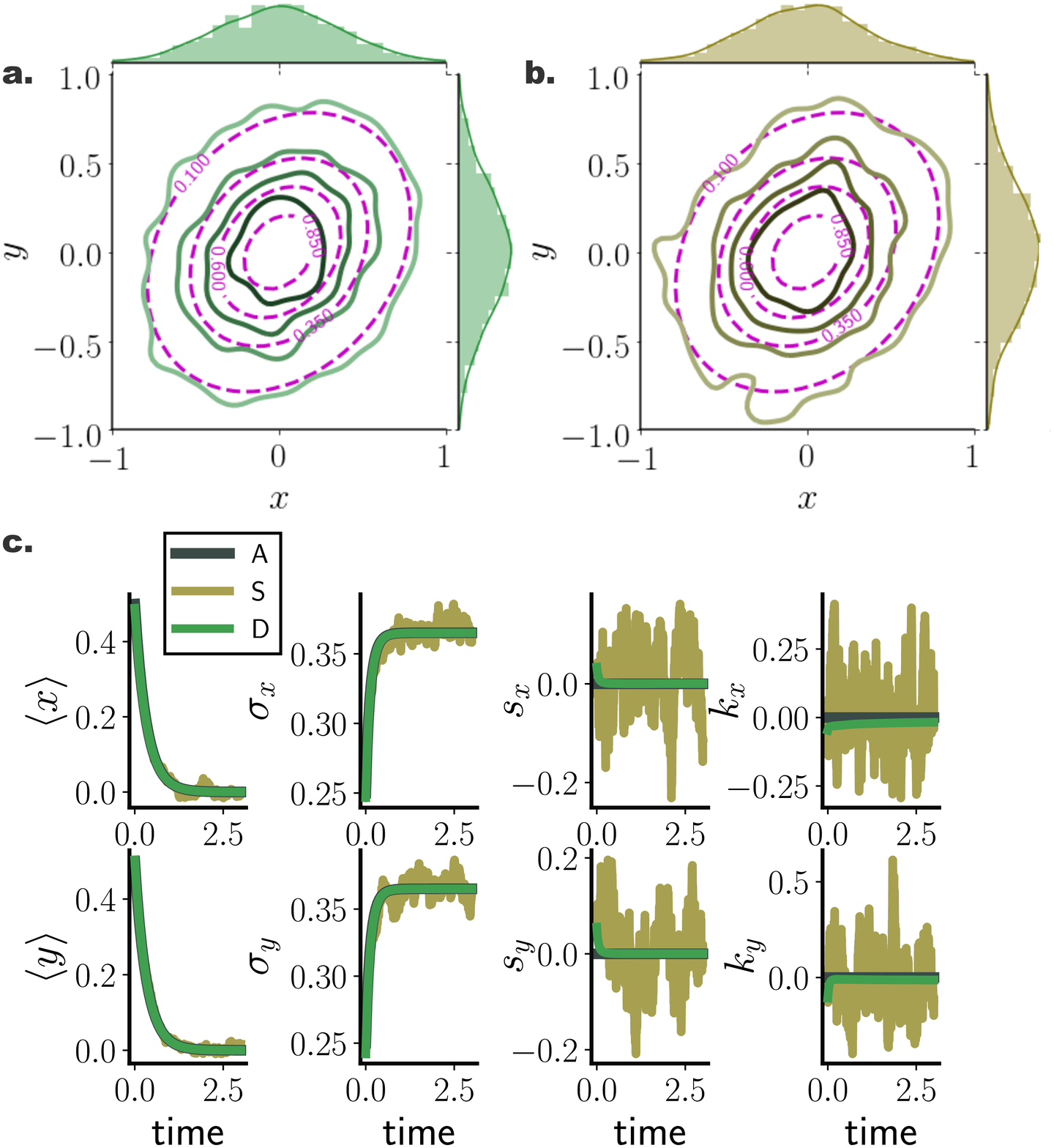}
  \caption{\textbf{Stationary and transient Fokker--Planck solutions computed with deterministic (green) and stochastic (brown) particle dynamics for a two dimensional Ornstein Uhlenbeck process.}\textbf{(a.,b.)} Estimated stationary PDFs arising from deterministic (${N=1000}$) (green), and  stochastic ($N=1000$) (brown) particle dynamics. Purple contours denote analytically calculated stationary distributions, while top and side histograms display marginal distributions for each dimension. \textbf{(c.)}  Temporal evolution of marginal statistics, mean $\langle x \rangle$, standard deviation $\sigma_x$, skewness $s_x$, and kurtosis $k_x$, for analytic solution ($A$), and for stochastic ($S$) and deterministic ($D$) particle systems comprising $N=1000$, with initial state distribution  $\protect\mathcal{N}\left(\left[\protect\begin{matrix}0.5 \\ 0.5 \protect\end{matrix}\right],\left[\protect\begin{matrix}0.05^2 & 0\\ 0 & 0.05^2\protect\end{matrix}\right]\right)$, for $M=100$ randomly selected inducing points employed in the gradient--log--density estimation. Deterministic particle simulations deliver smooth cumulant trajectories, as opposed to highly fluctuating stochastic particle cumulants.(Further parameter values: regularisation constant $\lambda=0.001$, and RBF kernel length scale $l$ estimated at every time point as two times the standard deviation of the state vector. Inducing point locations were selected randomly at each time step from a uniform distribution spanning the state space volume covered by the state vector.)}
  \label{fig:2d_OU}
\end{figure}

Likewise, we observed similar results when comparing only the stationary distributions, $ \mathcal{W}_1 (P_{\infty}^{A},P_{\infty}^N)$ (Fig.~\ref{fig:2d_OUb}(c.)). While for small particle number, the interacting particle system more accurately captured the underlying limiting distribution, for increasing particle number the accuracy of both approaches converged, with our method delivering consistently more reliable approximations among individual repetitions.

Moreover, densities evolved with our deterministic framework exhibited less fluctuating cumulant trajectories in time, compared to their stochastic counterparts (Fig.~\ref{fig:2d_OU}(c.)). In particular, even for limited particle number cumulants calculated over deterministically evolved particles progressed smoothly in time, while substantially more particles for the stochastic simulations were required for the same temporal cumulant smoothness. To quantify further the transient accuracy of Fokker--Planck solutions computed with our method, we compared the average transient discrepancy between the first two analytic cumulants ($m_t$ and $C_t$) to those estimated from the particles ($\hat{m}_t$ and $\hat{C}_t$), $\langle \|\hat{m}_t - m_t\|_2\rangle_t$ (Fig.~\ref{fig:2d_OU}(b.)) and $ \langle \| \hat{C}_t - C_t \|_F \rangle_t$ (Fig.~\ref{fig:2d_OU}(d.)). In line with our previous results, our deterministic framework delivered considerably more accurate transient cumulants, when compared to stochastic simulations, with more consistent results among individual realisations, denoted by smaller dispersion of average cumulant differences. (Notice the logarithmic y-axis scale in Fig.~\ref{fig:2d_OU}(b., d.). Error bars for the stochastic solutions were in fact larger than those for the deterministic solutions on a linear scale. )

Interestingly, the number of sparse points $M$ employed in the gradient--log--density estimation had only minor influence on  the quality of the solution (Fig.~\ref{fig:2d_OUb}(a., c.)). This hints to substantially low computational demands for obtaining accurate Fokker--Planck solutions, since our method is computationally limited by the inversion of the $M \times M$ matrix in Eq.~(\ref{eq:A}).

%%%%%%%%%%%%%%%%%%%%%%%%%%%%%%%%%%%%%%%%%%%%%%%%%%%%%%%%%%%%%%%%%%%%%%%%%%%%%
\subsection{Bi-stable nonlinear system with additive noise}

For nonlinear processes, since the transient solution of the FPE is analytically intractable, we compared the transient and stationary densities estimated by our method with those returned from stochastic simulations of $N^{\infty}=2650$ particles, and contrasted them against stochastic simulations with same particle number.

\begin{figure}[htbp]
 \centering 
  \includegraphics[width=0.9\textwidth]{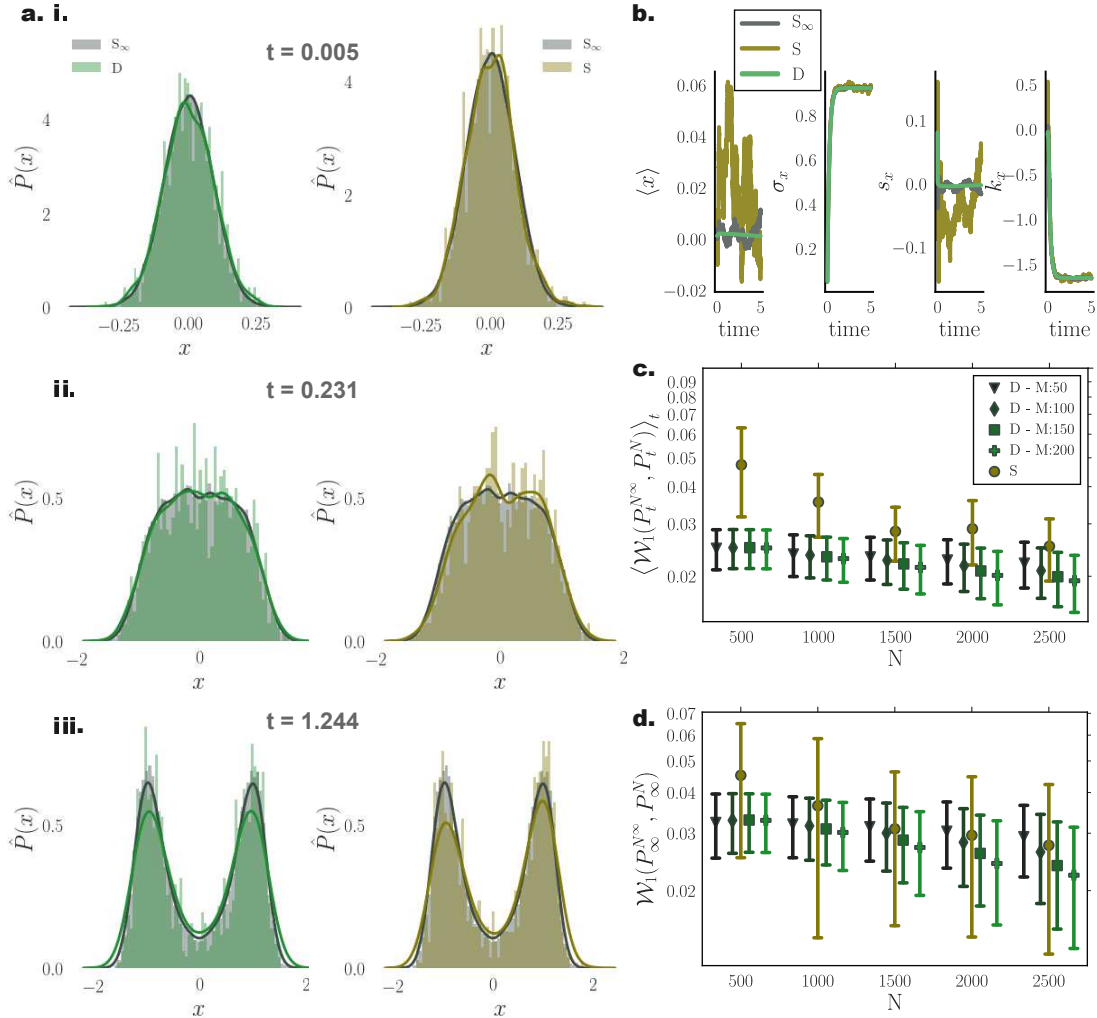}
  \caption{\textbf{Performance of deterministic (green) and stochastic (brown) $N$ particle solutions compared to $N^{\infty}$ (grey) stochastic particle densities for a nonlinear bi-stable process.} \textbf{(a.)} Instances of estimated pdfs arising from \textbf{(left)} stochastic ($N^{\infty}=26000$) (grey) and deterministic ($N=1000$) (green), and \textbf{(right)} stochastic ($N^{\infty}=2650$) (grey) and stochastic ($N=1000$) (brown) particle dynamics at times \textbf{(i.)} $t=0.005$, \textbf{(ii.)} $t=0.231$, and \textbf{(iii.)} $t=1.244$. \textbf{(b.)}  Temporal evolution of first four distribution cumulants, mean $\langle x \rangle$, standard deviation $\sigma_x$, skewness $s_x$, and kurtosis $k_x$, for stochastic ($S^{\infty}$ and $S$) and deterministic ($D$) systems comprising $N^{\infty} = 26000$, $N=1000$, with initial state distribution $\mathcal{N}(0,0.05^2)$, by employing $M=150$ inducing points in the gradient--log--density estimation. \textbf{(c.)} Mean, $\left< \mathcal{W}_1(P_t^{N^{\infty}},P_t^N) \right>_t$, and stationary, $\mathcal{W}_1 (P_{\infty}^{A},P_{\infty}^N)$,  1-Wasserstein distance, between $N^{\infty}=2650$ stochastic, and deterministic (D)/stochastic (S) simulations of $N$ particles (for different inducing point number $M$).  (Further parameter values: regularisation constant $\lambda=0.001$, Euler integration time step $dt=10^{-3}$, and RBF kernel length scale $l=0.5$. Inducing point locations were selected randomly at each time step from a uniform distribution spanning the state space volume covered by the state vector.) }
  \label{fig:1d_double_well}
\end{figure}

For a system with bi-modal stationary distribution (Appendix~\ref{ap:DW}), the resulting particle densities from our deterministic framework closely agreed with those arising from the stochastic system with $N^{\infty}=26000$ particles (Fig.~\ref{fig:1d_double_well}(a.)). In particular, deterministically evolved distributions respected the symmetry of the underlying double--well potential, while the stochastic system failed to accurately capture the potential symmetric structure Fig.~\ref{fig:1d_double_well}(a.iii.).

Systematic comparisons of the 1-Wasserstein distance between deterministic and stochastic $N$ particle simulations with the "$N^{\infty}$" stochastic simulation comprising $2650$ particles, revealed that our approach efficiently captured the underlying PDF already with $N=500$ particles (Fig.~\ref{fig:1d_double_well}(c.,d.)). For increasing particle number, the two systems converged to the "$N^{\infty}$" one. However, we observed a systematically increasing approximation accuracy delivered from the deterministic simulations compared to their stochastic counterparts. 

It is noteworthy, that on average deterministic simulations of $N=500$ particles conveyed a better approximation of the underlying transient PDF compared to stochastic simulations of $N=2500$ particles (Fig.~\ref{fig:1d_double_well}(c.)).
%Thus, although our approach might be more computationally demanding compared to stochastic simulations, BLAH [NEED TO MONITOR RUNTIME TO MAKE PRECISE STATEMENT]

Interestingly, for small particle number, the number of employed inducing points $M$ did not to influence significantly the accuracy of the approximated solution. However for increasing particle number, enlarging the set of inducing points contributed to more accurate approximation of Fokker--Planck equation solutions, with the trade off of additional computational cost.

Similar to the Ornstein Uhlenbeck process (Section~\ref{sec:2DOU}), comparing cumulant trajectories computed from both the deterministic and stochastic particle systems revealed less fluctuating cumulant evolution for densities evolved with our deterministic framework also in this nonlinear setting (Fig.~\ref{fig:1d_double_well}(b.)).

%%%%%%%%%%%%%%%%%%%%%%%%%%%%%%%%%%%%%%%%%%%%%%%%%%%%%%%%%%%%%%%%%%%%%%%%
\subsection{Nonlinear system perturbed by multiplicative noise}

\begin{figure}[htbp]
  \centering
  \includegraphics[width=0.72\textwidth]{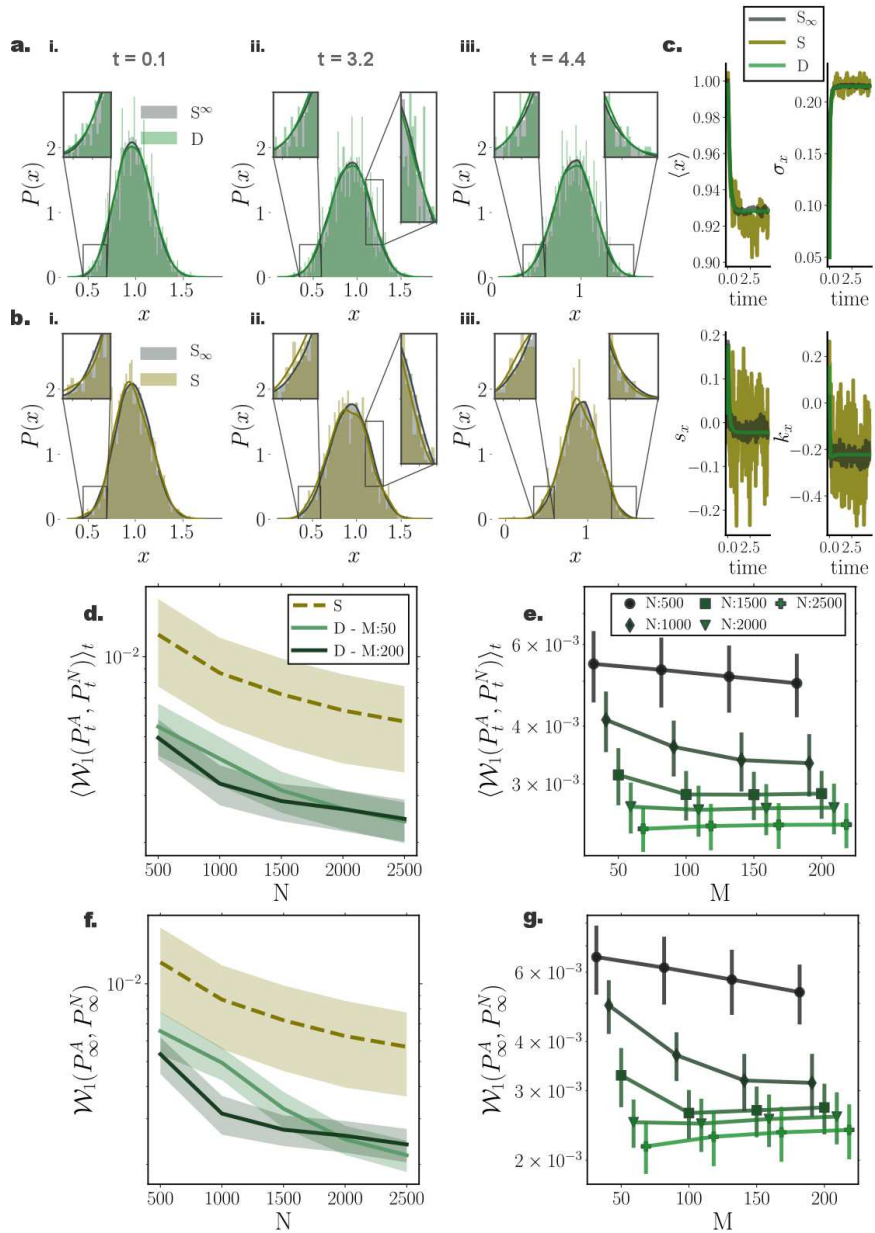}
  \caption{\textbf{Accuracy of Fokker--Planck solutions for a nonlinear system perturbed with state dependent noise.} \textbf{(a.)} Instances of $N=1000$ particle distributions resulting from deterministic (green) and \textbf{(b.)} stochastic (brown) simulations against stochastic particle distributions comprising $N^{\infty}=35000$ particles (grey) for \textbf{(i.)} $t = 0.1$, \textbf{(ii.)} $t = 3.2$, and \textbf{(iii.)} $t = 4.4$. Insets provide a closer view of details of distribution for visual clarity. Distributions resulting from deterministic particle simulations closer agree with underlying distribution for all three instances. \textbf{(c.)} Temporal evolution of first four cumulants for the three particle systems (grey: $S_{\infty}$ - stochastic with $N^{\infty}=35000$ particles, brown: $S$ - stochastic with $N=1000$ particles, and green: $D$ - deterministic with $N=1000$ particles). Deterministically evolved distributions result in smooth cumulant trajectories. \textbf{(d., e.)} Temporal average and \textbf{(f., g.)} stationary 1-Wasserstein distance between distributions mediated through stochastic simulations of $N^{\infty}=35000$, and through deterministic (green) and stochastic (brown) simulations of $N$ particles against particle number $N$. Shaded regions and error bars denote one standard deviation among $20$ independent repetitions. Different green hues designate different inducing point number $M$ employed in the gradient--log--density estimation. (Further parameter values: regularisation constant $\lambda=0.001$, Euler integration time step $dt=10^{-3}$, and RBF kernel length scale $l=0.25$. Inducing points were arranged on a regular grid spanning the instantaneous state space volume captured by the state vector.)
   }
  \label{fig:state_dependent}
\end{figure}

To asses the accuracy of our framework on general diffusion processes perturbed by state dependent (multiplicative) noise, we simulated a bi-stable system with dynamics governed by Eq.~(\ref{eq:DW}) with diffusion function $D(x)=sin^2(x)$ according to Eq.~(\ref{eq:statedep}). Also in this setting, deterministic particle distributions delivered a closer approximation of the underlying density, when compared to direct stochastic simulations. In particular, we found that in this setting, deterministically evolved distributions captured more accurately the tails of the underlying distribution, mediated here by stochastic simulations of $N^{\infty}=35000$ particles (Fig.~\ref{fig:state_dependent}(a.,b.)).

Similar to the previously examined settings, the deterministic framework delivered more reliable and smooth trajectories for the marginal statistics of the underlying distribution (Fig.~\ref{fig:state_dependent}(c.)).

Comparing the temporal average and stationary 1-Wasserstein distance (Fig.~\ref{fig:state_dependent}(d.,f.)) between the optimal stochastic distributions and the deterministic and stochastic particle distributions of size $N$, we found that the deterministic system delivered consistently more accurate approximations, as portrayed by smaller 1-Wasserstein distances.

Interestingly, we found that for deterministic particle simulations, the number of employed sparse points in the gradient--log--density estimation mediated a moderate approximation improvement for small system sizes, while for systems comprising more than $N=2000$ particles, the number of sparse points had minimal or no influence on the accuracy of the resulting distribution (Fig.~\ref{fig:state_dependent}(e.,g.)).

%%%%%%%%%%%%%%%%%%%%%%%%%%%%%%%%%%%%%%%%%%%%%%%%%%%%%%%%%%%%%%%%%%%%%

\subsection{Performance in higher dimensions}
\begin{figure}[htbp]
  \centering
  \includegraphics[width=1\textwidth]{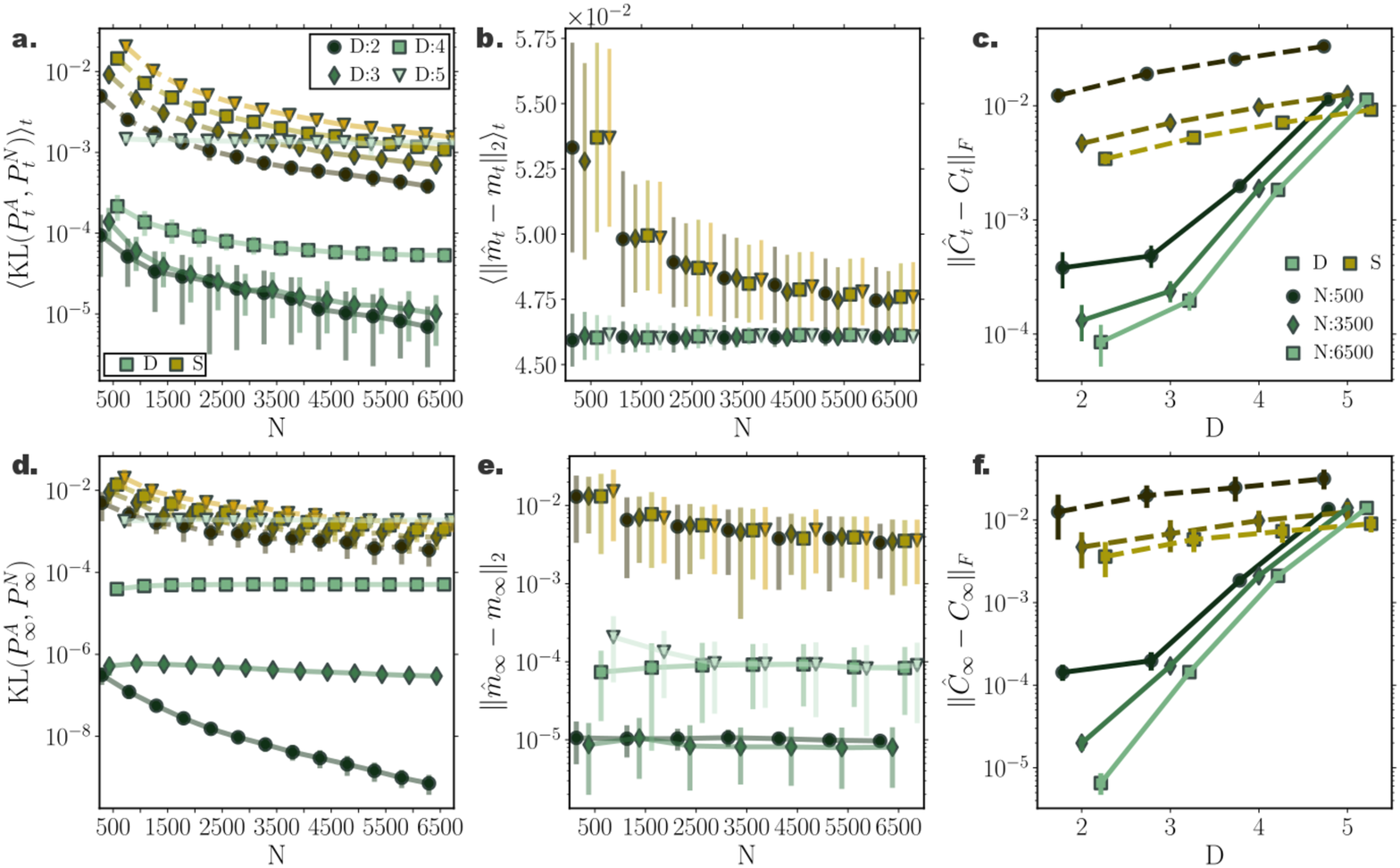}
  \caption{\textbf{Accuracy of Fokker-Planck solutions for multi-dimensional Ornstein--Uhlenbeck processes.} Comparison of deterministic particle Fokker--Planck solutions with stochastic particle systems and analytic solutions for multi-dimensional Ornstein--Uhlenbeck process of D=\{2,3,4,5\} dimensions. \textbf{(a.)} Time averaged and \textbf{(d.)} stationary Kullback--Leibler (KL) divergence between simulated particle solutions (black/green: deterministic, brown/yellow: stochastic) and analytic solutions for different dimensions. Deterministic particle simulations outperform stochastic particle solutions even for increasing system dimensionality.
 \textbf{(b.)} Time averaged and \textbf{(e.)} stationary error between analytic, $m_t$, and sample mean, $\hat{m_t}$, for increasing particle number.  
 \textbf{(c.)}  Time averaged and  \textbf{(f.)} stationary discrepancy between simulated, $\hat{C_t}$, and analytic covariances, $C_t$, as captured by the Frobenius norm of the relevant covariance matrices difference. The accuracy of the estimated covariance decreases for increasing dimensionality. 
 (Further parameter values: number of inducing points $M=100$, regularisation constant $\lambda=0.001$, Euler integration time step $dt=10^{-3}$, and adaptive RBF kernel length scale $l$ calculated at every time step as two times the standard deviation of the state vector.
 Inducing point locations were selected randomly at each time step from a uniform distribution spanning the state space volume covered by the state vector.)\hfill  }
  \label{fig:Nd_OU}
\end{figure}

To quantify the scaling and performance of the proposed framework for increasing system dimension, we systematically compared simulated densities with analytically calculated ones for Ornstein--Uhlenbeck processes of dimension $D=\{2,3,4,5\}$ following the dynamics of Eq.~(\ref{eq:multiOU}). To evaluate simulated Fokker--Planck solutions we calculated Kullback--Leibler divergence between analytically evolved densities (Appendix~\ref{ap:SolOU}) and particle densities. We employed the closed form equation for estimating KL divergence between two Gaussian distributions (Appendix~\ref{sec:KL}) for empirically estimated mean, $\hat{m_t}$, and covariance, $\hat{C_t}$, for particle distributions.

For all dimensionalities, the deterministic particle solutions approximated transient and stationary densities remarkably accurately with Kullback--Leibler divergence between the simulated and analytically derived densities below $10^{-2}$ for all dimensions, both for transient and stationary solutions (Fig.~\ref{fig:Nd_OU}(a.,d.). In fact, the deterministic particle solutions delivered more precise approximations of the underlying densities compared to direct stochastic simulations of the same particle number. Remarkably, even for processes of dimension $D=5$ deterministically evolved solutions mediated through $N=500$ particles resulted in approximately same KL divergence of stochastic particle solutions of $N=6500$ particles. 

Our deterministic particle method delivered consistently better approximations of the mean of the underlying densities compared to stochastic particle simulations (Fig.~\ref{fig:Nd_OU}(b.,e.). Specifically, estimations of the stationary mean of the underlying distributions were more than two orders of magnitude accurate that their stochastically approximated counterparts already for small particle number (Fig.~\ref{fig:Nd_OU}(e.). 

Yet, the accuracy of our deterministic framework deteriorated for increasing dimension (Fig.~\ref{fig:Nd_OU}(a.,d.). More precisely, although for low dimensionalities the covariance matrices of the underlying densities were accurately captured by deterministically evolved particles, for increasing system dimension approximations of covariance matrices became progressively worse. Yet, even for systems of dimension $D=5$, covariance matrices computed from deterministically simulated solutions of $N=500$ particles were at the same order of magnitude as accurate as covariances delivered by stochastic particle simulations of size $N=6500$.

%%%%%%%%%%%%%%%%%%%%%%%%%%%%%%%%%%%%%%%%%%%%%%%%%%%%%%%%%%

\subsection{Second order Langevin systems}

\begin{figure}[htbp]
  \centering
  \includegraphics[width=0.95\textwidth]{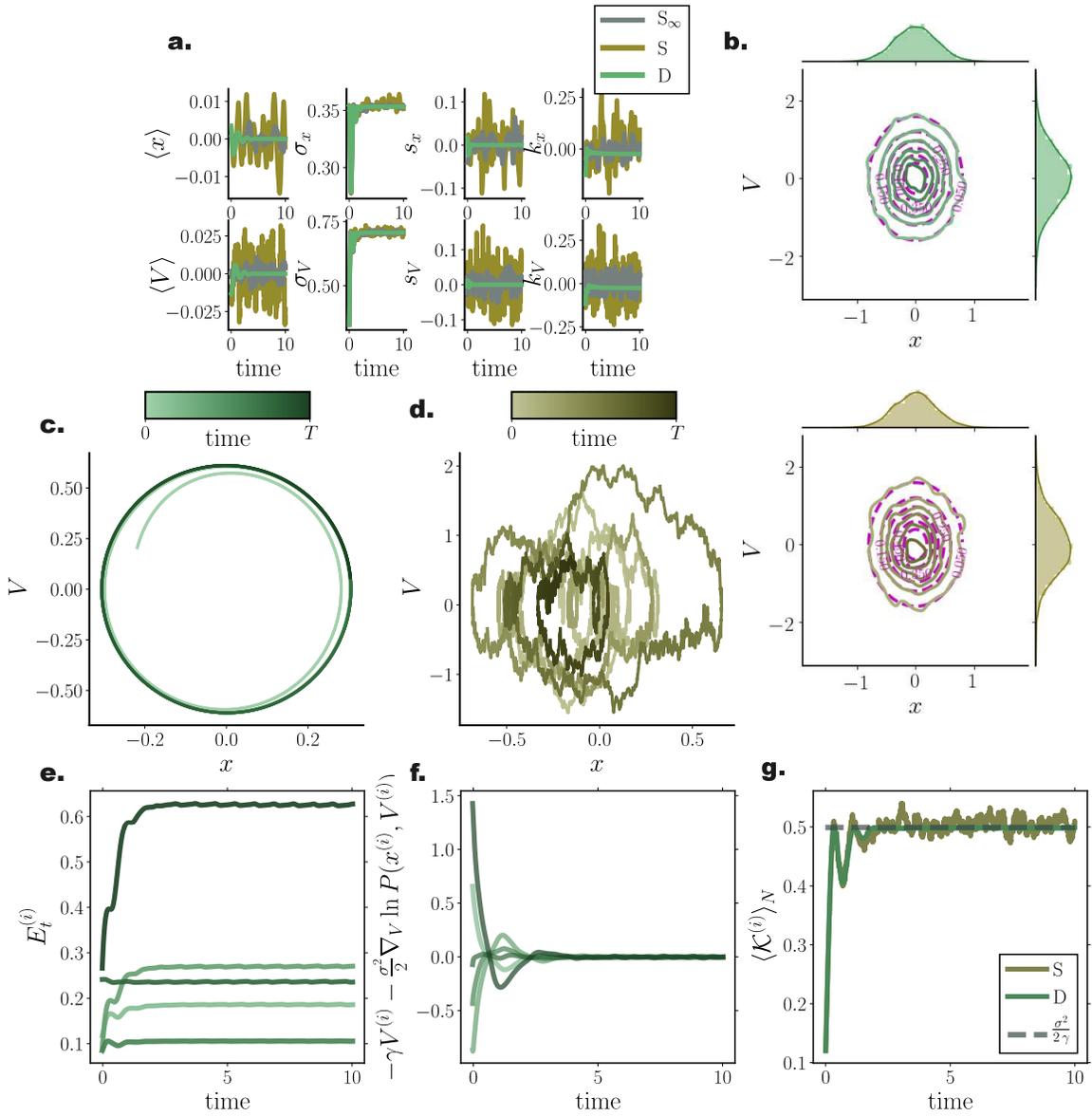}
  \caption{\textbf{Energy preservation for second order Langevin dynamics in a quadratic potential.} Comparison of deterministic particle Fokker--Planck solutions with stochastic particle systems for a harmonic oscillator \textbf{(a.)} First four cumulant temporal evolution for deterministic (green) and stochastic (brown) system.
  \textbf{(b.)} Stationary joint and marginal distributions for deterministic (green) and stochastic (brown) systems. Purple lines denote analytically derived stationary distributions.
    \textbf{(c., d.)} State space trajectory of a single particle for deterministic (green) and stochastic (brown) system. Color gradients denote time.
  \textbf{(e.)} Temporal evolution of individual particle energy $E^{(i)}_t$ for deterministic system for 5 particles.
 \textbf{(f.)} Difference between velocity and gradient--log--density term for individual particles. After the system reaches stationary state the particle velocity and GLD term cancel out.
 \textbf{(g.)} Ensemble average kinetic energy through time resorts to $\frac{\sigma^2}{2\,\gamma}$ (grey dashed line) after equilibrium is reached.
 (Further parameter values: regularisation constant $\lambda=0.001$, integration time step $dt=2\cdot 10^{-3}$, and adaptive RBF kernel length scale $l$ calculated at every time step as two times the standard deviation of the state vector. Number of inducing points $M=300$. Inducing point locations were selected randomly at each time step from a uniform distribution spanning the state space volume covered by the state vector.)  }
  \label{fig:Ham}
\end{figure}

\begin{figure}[htbp]
  \centering
  \includegraphics[width=1\textwidth]{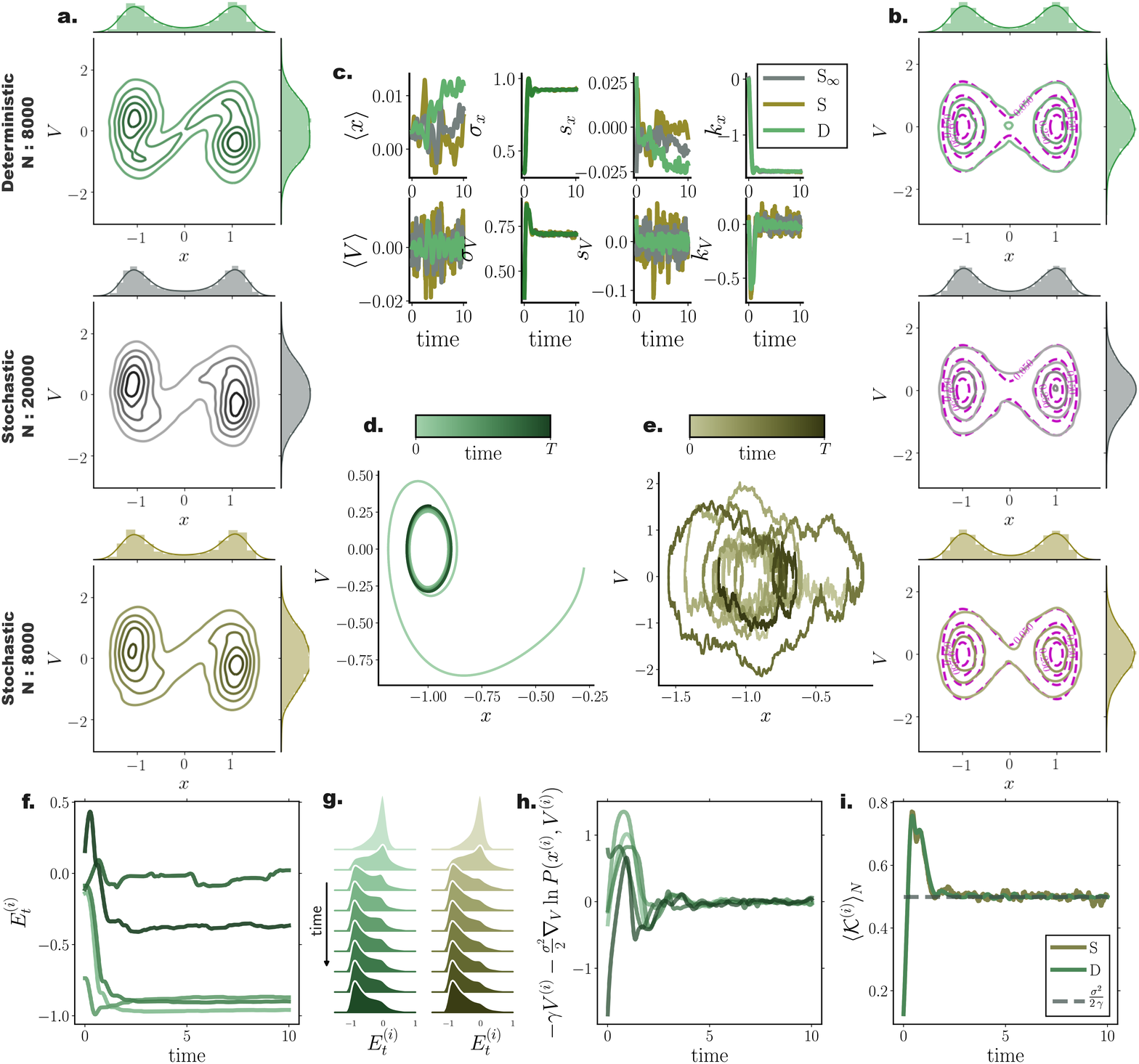}
  \caption{\textbf{Energy preservation for second order Langevin dynamics in a double well potential.} Comparison of deterministic particle Fokker--Planck solutions with stochastic particle systems for a bistable process. \textbf{(a., b.)} Joint and marginal distributions of system states mediated by $N=8000$ particles evolved with our framework (green) and with direct stochastic simulations comprising $N^{\infty}=20000$ (grey) and $N=8000$ (brown) particles at \textbf{(a.)} $t=0.6$, and \textbf{(b.)} $t=10$. Purple lines denote the analytically derived stationary density. \textbf{(c.)} First four cumulant temporal evolution for deterministic (green) and stochastic (brown) system.
   \textbf{(d.)} State space trajectory of a single particle for deterministic and \textbf{(e.)} stochastic system. Color gradients denote time.
  \textbf{(f.)} Temporal evolution of individual particle energy $E^{(i)}_t$ for deterministic system for $5$ particles.
  \textbf{(g.)} Temporal evolution of distribution of particle energies $E^{(i)}_t$ for deterministic (green) and stochastic (brown) system.
 \textbf{(h.)} Difference between velocity and gradient log density term for individual particles.
  \textbf{(i.)} Ensemble average kinetic energy through time resorts to $\frac{\sigma^2}{2\,\gamma}$ (grey dashed line) after equilibrium is reached.
 (Further parameter values: regularisation constant $\lambda=0.001$, integration time step $dt=2\cdot 10^{-3}$ and adaptive RBF kernel length scale $l$ calculated at every time step as two times the standard deviation of the state vector. Number of inducing points $M=300$. Inducing point locations were selected randomly at each time step from a uniform distribution spanning the state space volume covered by the state vector. ) }
  \label{fig:HamNon}
\end{figure}

To demonstrate the performance of our framework for simulating solutions of the FPEs for second order Langevin systems as described in Section~\ref{sec:Langevin}, we incorporated our method in a symplectic Verlet integrator (Eq.(~\ref{eq:symplectic}-~\ref{eq:symplectic2})) simulating the second order dynamics captured by Eq.~(\ref{Kramers_det}) for a linear $f(x)=-4\,x$ and a nonlinear, $f(x)=-4\,x^3+4\,x$, drift function (Eq.~(\ref{eq:symplectic})), and compared the results with stochastic simulations integrated by a semi-symplectic framework~\cite{milstein2007computing}.
In agreement with previous results, cumulant trajectories evolved smoother in time for deterministic particle simulations when compared to their stochastic counterparts (Fig.~\ref{fig:Ham}(a.) and Fig.~\ref{fig:HamNon}(c.)).
Stationary densities closely matched analytically derived ones (see Eq.~(\ref{eq:Lang_station})) (purple contour lines in Fig.~\ref{fig:Ham}(b.) and Fig.~\ref{fig:HamNon}(b.)), while transient densities captured the fine details of simulated stochastic particle densities comprising $N^{\infty}=20000$ (Fig.~\ref{fig:HamNon}(a.)).

Furthermore, the symplectic integration contributed to the preservation of energy levels for each particle, after the system reached equilibrium (Fig.~\ref{fig:Ham}(e.) and Fig.~\ref{fig:HamNon}(f.)), which was also evident when observing individual particle trajectories in the state space (Fig.~\ref{fig:Ham}(c., d.) and Fig.~\ref{fig:HamNon}(d., e.)).

As already conveyed in Section~\ref{sec:Langevin}, the velocity term and the gradient--log--density term canceled out in the long time limit (Fig.~\ref{fig:Ham}(f.) and Fig.~\ref{fig:HamNon}(g.))  for each particle individually, while the average kinetic energy in equilibrium exactly resorted to the value dictated by the fluctuation--dissipation relation and the equipartition of energy property, i.e.  $\langle \mathcal{K}^{(i)} \rangle_N=\frac{\sigma^2}{2\,\gamma}$ (Fig.~\ref{fig:Ham}(g.) and Fig.~\ref{fig:HamNon}(h.)).

%Moreover, the average energy of the system exhibited strong temporal fluctuations for the stochastic particle densities, while our framework delivers rather smooth average energy trajectories (Fig.~\ref{fig:Ham}(h.) and Fig.~\ref{fig:HamNon}(h.)). 

\subsection{Nonconservative chaotic system with additive noise (Lorenz63) }

\begin{figure}[htbp]
  \centering
  \includegraphics[width=0.74\textwidth]{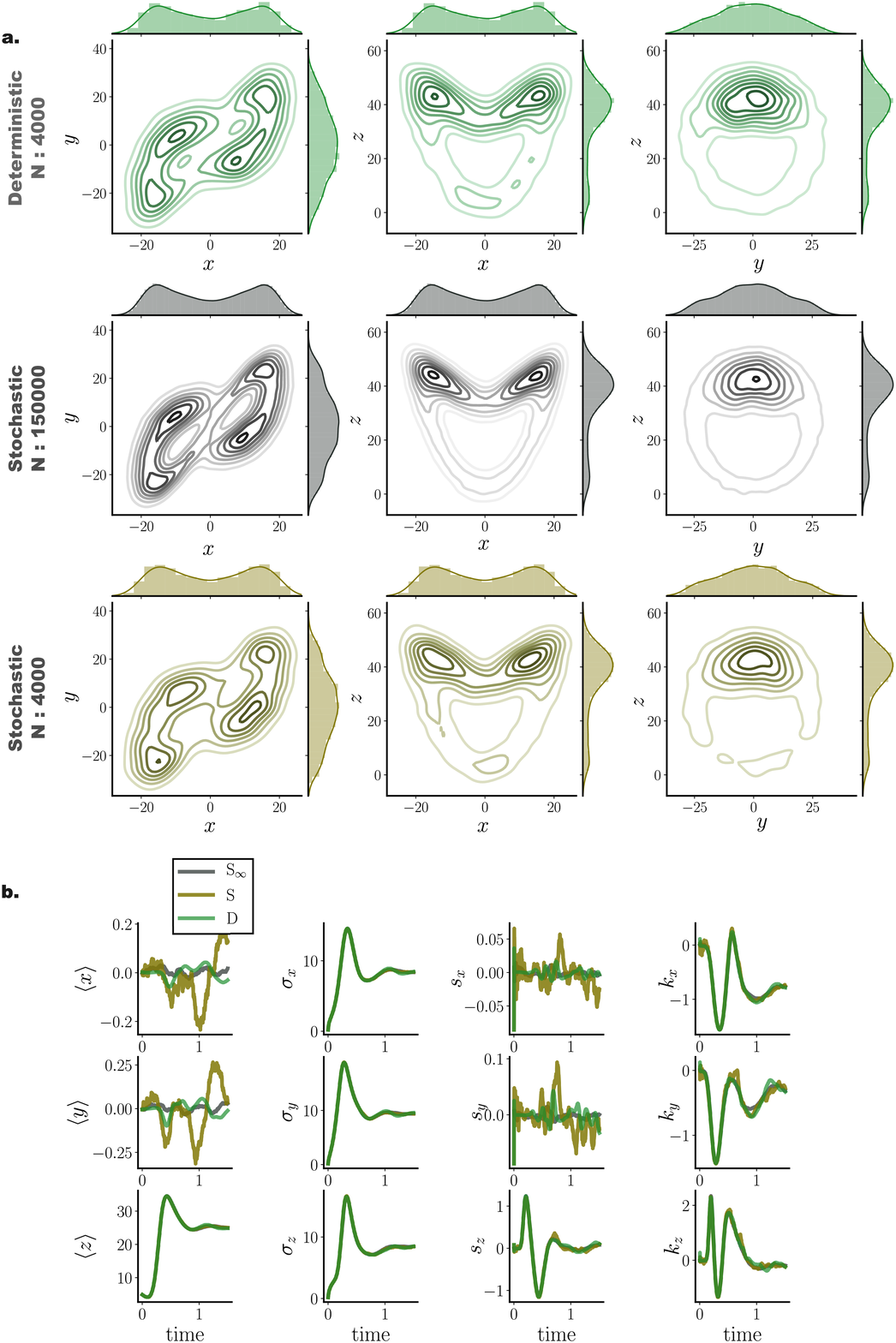}
  \caption{\textbf{Deterministic (green) and stochastic (brown) Fokker--Planck particle solutions for a three dimensional Lorenz63 system in the chaotic regime perturbed by additive Gaussian noise.} \textbf{(a.)} Joint and marginal distributions of system states mediated by $N=4000$ particles evolved with our framework (green) and with direct stochastic simulations comprising $N=150000$ (grey) and $N=4000$ (brown) particles at $t=0.4$. %Joint distributions computed through deterministic particle trajectories capture finer details of the underlying distribution (grey) compared to stochastic simulations (brown). 
  \textbf{(b.)} Cumulant trajectories for the three particle systems. Cumulants derived from deterministic particle simulations (green) closer match cumulant evolution of the  underlying distribution (grey) compared to stochastic simulations (brown). (Further parameter values: regularisation constant $\lambda = 0.001$, Euler integration time step $dt=10^{-3}$, adaptive RBF kernel length scale $l$ calculated at every time step as two times the standard deviation of the state vector. Number of inducing points: $M=200$. Inducing point locations were selected randomly at each time step from a uniform distribution spanning the state space volume covered by the state vector.)  }
  \label{fig:Lorenz}
\end{figure}

As a final assessment of our framework for simulating accurate solutions of Fokker--Planck equations, we employed a Lorenz63 model with parameters rendering the dynamics chaotic, perturbed by moderate additive Gaussian noise (Eq.~(\ref{eq:Lorenz})). By comparing stochastic simulations of ${N^{\infty}=150000}$ particles and deterministic and stochastic simulations of $N=4000$ particles (Fig.~\ref{fig:Lorenz}), we observed that the deterministic framework captured more precisely finer details of the underlying distribution (Fig.~\ref{fig:Lorenz}(a.)), represented here by the $N^{\infty}$ stochastic simulation. While both stochastic and deterministic simulations capture the overall butterfly profile of the Lorenz attractor, the deterministic system delivered indeed a closer match to the underlying distribution.

Similar to the previously examined models, cumulant trajectories computed from deterministically evolved particles show closer agreement with those computed from the $N^{\infty}$ stochastic system, compared to the stochastic system comprising $N$ particles (Fig.~\ref{fig:Lorenz}(b.)). In particular, cumulants for the $x$ and $y$ states exhibited high temporal fluctuations when computed from stochastically evolved distributions, while our framework conveyed more accurate cumulant trajectories, closer to those delivered by the $N^{\infty}$ stochastic system.

\section{Discussion and Outlook} \label{sec:outlook}

We presented a particle method for simulating solutions of FPEs governing the temporal evolution of the probability density for 
stochastic dynamical systems of the diffusion type.
By reframing the FPE in a Liouville form,
we obtained an effective dynamics in terms of independent deterministic particle trajectories. Unfortunately, this formulation requires the knowledge of the gradient of the logarithm of the instantaneous probability density of the system state, which is the quantity we try to compute. We circumvented this complication by introducing statistical estimators for the gradient--log--density based on a variational formulation. To combine high flexibility of estimators with computational 
efficiency, we employed kernel based estimation together with an additional sparse approximation. For the case of equilibrium systems, we related our framework to 
Stein Variational Gradient Descent, a particle based dynamics to approximate 
the stationary density, and to a geometric formulation of Fokker--Planck dynamics.
We further discussed extensions of our method to settings with multiplicative noise and to second order Langevin dynamics.

 To demonstrate the performance of our framework, we provided detailed tests
 and comparisons with stochastic simulations and analytic solutions (when possible). We demonstrated the accuracy of our method on conservative and non-conservative model systems 
 with different dimensionalities.  In particular, we found, that our framework outperforms stochastic simulations both in linear and nonlinear settings, by delivering more accurate densities for small particle number when the dimensionality is small enough. For increasing particle number, the accuracy of both approaches converges. Yet, our deterministic framework delivered \emph{consistently} results with smaller variance among individual repetitions.
 Furthermore, we showed that our method, even for small particle numbers,
 exhibits low order cumulant trajectories with significantly less temporal fluctuations  when compared against to stochastic simulations of the same particle number. 
 
 We envisage several ways to improve and extend our method.
There is room for improvement by optimising hyper parameters of our algorithm such as 
inducing point position and kernel length scale.
Current grid based and uniform random selection of inducing point position may contribute to 
the deterioration of solution accuracy in higher dimensions. Other methods, 
such as subsampling or clustering of particle positions may lead to further improvements. 
On the other hand, a hyper parameter update may not be at all necessary 
at each time step in certain settings, such that a further speedup of our algorithm could be achieved.

The implementation of our method depends on the function class chosen to represent the estimator. In this paper we have focused on linear representations, leading to simple 
closed form expressions. It would be interesting to see if 
other, nonlinear parametric models, such as neural networks,  
(see e.g.~\cite{deepnn}) could be employed to represent estimators. 
While, in this setting, there would be no closed form solutions, the small changes 
in estimates between successive time steps, suggest that
only a few updates of numerical optimisation may be necessary 
at each step. Moreover, the ability of neural networks 
to automatically learn relevant features from data might help to improve 
performance for higher dimensional problems when 
particle motion is typically restricted on lower dimensional submanifolds. 

From a theoretical point of view, rigorous results on the 
accuracy of the particle approximation would be important. These would 
depend on the speed of convergence of estimators towards exact gradients 
of log--densities. However, to obtain such results may not be easy.
While rates of convergence for kernel based estimators 
have been studied in the literature, the methods for proofs usually 
rely on the independence of samples and would not necessarily apply to the 
case of interacting particles.  

We have so far addressed only the forward simulation of FPEs. However, preliminary results indicate that related techniques may be applied to particle based simulations for smoothing (forward--backward) and related control problems for diffusion processes~\cite{reich2015probabilistic}. Such problems involve computations of an effective, controlled drift function in terms of gradient--log--densities. We defer further details and discussions on subsequent publications on the topic.

Taken together, the main advantage of our framework is its minimal requirement in simulated particle trajectories for attaining reliable Fokker--Planck solutions with smoothly evolving transient statistics. Moreover, our proposed method is nearly effortless to set up when compared to classical grid based FPE solvers, while it delivers more reliable results than direct stochastic simulations.

\centerline{--------------------------------------------}

%%%%%%%%%%%%%%%%%%%%%%%%%%%%%%%%%%%%%%%%%%

%%%%%%%%%%%%%%%%%%%%%%%%%%%%%%%%%%%%%%%%%%
%% optional
%\supplementary{The following are available online at \linksupplementary{s1}, Figure S1: title, Table S1: title, Video S1: title.}


%%%%%%%%%%%%%%%%%%%%%%%%%%%%%%%%%%%%%%%%%%
\authorcontributions{Conceptualization, S.R. and M.O.; methodology, D.M. and M.O.; software, D.M.; validation, D.M. and M.O.; formal analysis, D.M. and M.O.; investigation, D.M.; resources, M.O.; data curation, D.M.; writing--original draft preparation, D.M. and M.O.; writing--review and editing, D.M., S.R and M.O.; visualization, D.M.; supervision, M.O.; project administration, M.O.; funding acquisition, S.R. and M.O.}

%%%%%%%%%%%%%%%%%%%%%%%%%%%%%%%%%%%%%%%%%%
\funding{This research has been partially funded by Deutsche Forschungsgemeinschaft (DFG) - SFB1294/1 - 318763901.}

%%%%%%%%%%%%%%%%%%%%%%%%%%%%%%%%%%%%%%%%%%
%\acknowledgments{This research has been partially funded by Deutsche Forschungsgemeinschaft (DFG) through grants CRC 1294 ‘Data Assimilation’ (project A06).}

%%%%%%%%%%%%%%%%%%%%%%%%%%%%%%%%%%%%%%%%%%
\conflictsofinterest{The authors declare no conflict of interest. The funders had no role in the design of the study; in the collection, analyses, or interpretation of data; in the writing of the manuscript, or in the decision to publish the results.} 

%=====================================
% References, variant A: internal bibliography
%=====================================
\reftitle{References}
\externalbibliography{yes}
\bibliography{references.bib}
%\printbibliography
%% optional
\appendixtitles{yes} %Leave argument "no" if all appendix headings stay EMPTY (then no dot is printed after "Appendix A"). If the appendix sections contain a heading then change the argument to "yes".
\appendix
\section{Simulated systems}

\subsection{Two dimensional Ornstein-Uhlenbeck process} \label{ap:OU2}

For comparing Fokker-Planck solutions computed with our approach with solutions derived from stochastic simulations, we considered the two dimensional Ornstein-Uhlenbeck process captured by the following equations

\begin{align} \label{eq:2DOU}
    dX_t &= \left( - 4 X_t + Y_t \right) dt + \sigma \, dB_1\\
    dY_t &= \left( - 4 Y_t + X_t \right) dt + \sigma \, dB_2,
\end{align}
where the related potential is $U(x,y) = 2x^2 - x\,y +2y^2$.
Simulation time was set to $T=3$ with Euler–Maruyama integration step $dt = 10^{-3}$. For estimating the instantaneous gradient log density we employed $M=\{50,100,150,200 \}$ inducing points, randomly selected at every time point from a uniform distribution spanning the state space volume covered by the particles at the current time point.

\subsection{Bistable nonlinear system} \label{ap:DW}

For testing our framework on nonlinear settings, we simulated 
\begin{align}\label{eq:DW}
    dX_t &= \left( -4 {X_t}^3+4 X_t\right) dt +\,D(X_t)^{\frac{1}{2}} dB_t,
\end{align}
with $D(x) =1 $ for evaluating solutions with additive Gaussian noise, and with $D(x)=sin^2(x)$ for multiplicative noise FP solutions. The associated potential reads $U(x) = x^4- 2x^2$.

\subsection{Multi-dimensional Ornstein-Uhlenbeck processes}\label{ap:OUD}

For quantifying the scaling of our method for increasing system dimension, we  simulated systems of dimensionality $D=\{2,3,4,5\}$ according to the following equation

\begin{equation} \label{eq:multiOU}
 d{X_{(i)}}_t = \left( - 4 {X_{(i)}}_t + \sum^d_{j=1 , i\neq j} \frac{1}{2}{X_{(j)}}_t \right) dt + dB_{(i)},
\end{equation}
for $d \in D$.
Simulation time was determined by the time required for analytic mean $m_t$ to converge to its stationary solution within $\tilde{\epsilon}$ precision $\tilde{\epsilon} = 10^{-5}$, while  the integration step was set to $dt = 10^{-3}$.  

\subsection{Second order Langevin dynamics}

For demonstrating the energy preservation properties of our method for second order Langevin dynamics, we incorporated our framework into a Verlet symplectic integration scheme (Eq.~(\ref{eq:symplectic})), and compared the results with stochastic simulations integrated according to a semi-symplectic scheme~\cite{milstein2007computing}.

We consider a system with dynamics for positions $X$ and velocities $V$ captured by

\begin{align} \label{eq:app_lang}
dX &= V dt \\ 
dV &= \left(- \gamma V + f(X)\right) dt + \sigma dB_t\;,
\end{align} 
where the velocity change (acceleration) is the sum of a deterministic drift $f$, a velocity dependent damping $- \gamma V$, and a stochastic noise term $\sigma dB_t$. 

In conservative settings the drift comes as the gradient of a potential $f(x)=-\nabla U(x)$. Here we used a quadratic (harmonic) potential $U(x)=2x^2$ and a double-well potential $U(x) = x^4- 2x^2$.

In equilibrium, the Fokker--Planck solution is the Maxwell--Boltzmann distribution, i.e. 
\begin{equation} \label{eq:Lang_station}
 p_\infty(X,V) = \frac{1}{Z}e^{-  \beta H(X,V)  }= \frac{1}{Z}e^{-  \beta\left(\frac{\|V\|^2}{2} + U(X)\right)  },
\end{equation}
with partition function $Z = \displaystyle \int{ e^{-  \beta\left(\frac{\|V\|^2}{2} + U(X)\right)  } }dx dv$.

We may compute the energy of each particle at each time point as the sum of its kinetic and potential energies
\begin{equation} \label{eq:particle_energy}
    E_t^{(i)} = \frac{1}{2} {V^{(i)}}^2 + U(X^{(i)}).
\end{equation}
Here the superscripts denote individual particles.
After the system has reached equilibrium, energy levels per particle are expected to remain constant.

From the equipartition of energy and the fluctuation--dissipation relation, in the long time limit the average kinetic energy of the system is expected to resort to 
\begin{equation} \label{eq:kin}
    \lim_{t \to \infty} \langle \mathcal{K}\rangle = \lim_{t \to \infty} \frac{1}{2} \langle V^2 \rangle = \frac{\sigma^2}{2\,\gamma}.
\end{equation}

Symplectic integration~\cite{leimkuhler2004simulating} of Eq.~(\ref{Kramers_det}) follows the equations
\begin{align} \label{eq:symplectic}
    V_{n+\frac{1}{2}} &= V_n + \frac{dt}{2}\left( -\gamma V_n + f(X_n) -\frac{\sigma^2}{2} \nabla_v \ln p_t(X_n,V_n)   \right)\\
    X_{n+1} &= X_n+ dt\, V_{n+\frac{1}{2}}\\
    V_{n+1} &= V_{n+\frac{1}{2}} + \frac{dt}{2}\left( -\gamma V_{n+\frac{1}{2}} + f(X_{n+1}) -\frac{\sigma^2}{2} \nabla_v \ln p_t(X_{n+1},V_{n+\frac{1}{2}})   \right),\label{eq:symplectic2}
\end{align}
where $n$ denotes a single integration step.

\subsection{Lorenz63}
For simulating trajectories of the noisy Lorenz63 system we employed the following equations

\begin{align} \label{eq:Lorenz}
    dx_t &= \sigma (y-x) dt +\sigma dW_x\\
    dy_t &= \left( x (\rho -z)-y  \right) dt + \sigma dB_y\\
    dz_t &= \left( x\,y -\beta z \right) dt + \sigma dB_z,
\end{align}
with parameters $\sigma=10$, $\rho = 28$, and $\beta = \frac{8}{3}$, that render the deterministic dynamics chaotic~\cite{lorenz1963deterministic}, employing moderate additive Gaussian noise.

\section{Computing central moment trajectories for linear processes}\label{ap:SolOU}

For a linear process
\begin{equation}dX_t = A\,X_t dt +\sigma dB,
\end{equation}
the joint density of the state vector $X$ remains Gaussian
for all times when the initial density is Gaussian. The
mean vector $m$ and covariance matrix $C$ may be computed by solving the
ODE system
\begin{eqnarray}
\frac{dm}{dt} = A m \\
\frac{dC}{dt} = AC + CA^\top + \sigma^2 I. 
\end{eqnarray}

\section{Kullback--Leibler divergence for Gaussian distributions} \label{sec:KL}

We calculated the KL divergence between the theoretical and simulated distributions with

    \begin{equation}
      KL\left( P_1 || P_2 \right) = \frac{1}{2}\left( \log(|\Sigma_2|/|\Sigma_1|)- d + Tr( \Sigma_2^{-1} \Sigma_1) + (\mu_2-\mu_1)^T  \Sigma_2^{-1}  (\mu_2-\mu_1) \right),
    \end{equation}

where $P_x \sim \mathcal{N} \left(\mu_x,\Sigma_x  \right)$.

\section{Wasserstein distance} \label{sec:Wasser}

We employed the 1-Wasserstein distance \cite{villani2008optimal} as a distance metric for comparing pairs of empirical distributions. 

For two distributions $P$ and $Q$, we denote with $\mathcal{J}(P,Q)$ all joint distributions $J$ for a pair of random variables $(X,Y)$ with marginals $P$ and $Q$. Then the \emph{Wasserstein distance} between these distributions reads
\begin{equation}
    \mathcal{W}_p(P,Q) = \left( \text{inf}_{J \in \mathcal{J}(P,Q)} \int \| x-y \|^p dJ(x,y)     \right)^{\frac{1}{p}},
\end{equation}
where for the $1$-Wasserstein distances (used in the present manuscript) $p=1$. 

Interestingly, the Wasserstein distance between two one dimensional distributions $P$ and $Q$ obtains a closed form solution
\begin{equation}
  \mathcal{W}_p(P,Q) = \left( \int^1_0  \| F_P^{-1}(\tau) - F_Q^{-1} (\tau) \|^p d\tau \right)^{1/p},
\end{equation}
with $F_P$ and $F_Q$ indicating the cumulative distribution functions of P and Q.

Moreover, for one dimensional empirical distributions $P$ and $Q$ with samples of same size $\{ X_{i} \}^n_{i=1}$ and $\{ Y_{i} \}^n_{i=1}$, the Wasserstein distance simplifies into computation of differences of order statistics

\begin{equation}
    \mathcal{W}_p(P,Q) = \left(  \sum^n_{i=1} \| X_{(i)} - Y_{(i)}  \|^p\right)^{\frac{1}{p}},
\end{equation}
where $X_{(i)}$ and $Y_{(i)}$ indicates the $i$-th order statistic of the sample $\{ X_{i} \}^n_{i=1}$ and $\{ Y_{i} \}^n_{i=1}$, i.e. ${X_{(1)} \leq X_{(2)} \leq \dots \leq X_{(n)}}$ and ${Y_{(1)} \leq Y_{(2)} \leq \dots \leq Y_{(n)}}$.

\section{Influence of hyperparameter values on the performance of the Gradient--Log--Density estimator}

To determine the influence of the hyperparameter values on the performance of the gradient--log--density estimator, we systematically evaluated the approximation error of our estimator for $N=1000$ samples of a one dimensional log--normal distribution with mean $\mu=0$ and standard deviation $\sigma=0.5$ for $20$ independent realisations.

We quantified the approximation error as the average error between the analytically calculated and predicted gradient-log-density on each sample, i.e.
\begin{equation}
    \text{Approximation error} = \frac{1}{N} \sum^N_{i=1} \| \nabla \ln p(x_i) - \widehat{(\nabla \ln p(x_i)) } \| ,
\end{equation}

where the analytically calculated gradient-log-density was determined as ${\nabla \ln p(x) = \frac{\mu-\sigma^2-\ln(x)}{\sigma^2\,x}}$.

\begin{figure}[htbp]
  \centering
  \includegraphics[width=1.1\textwidth]{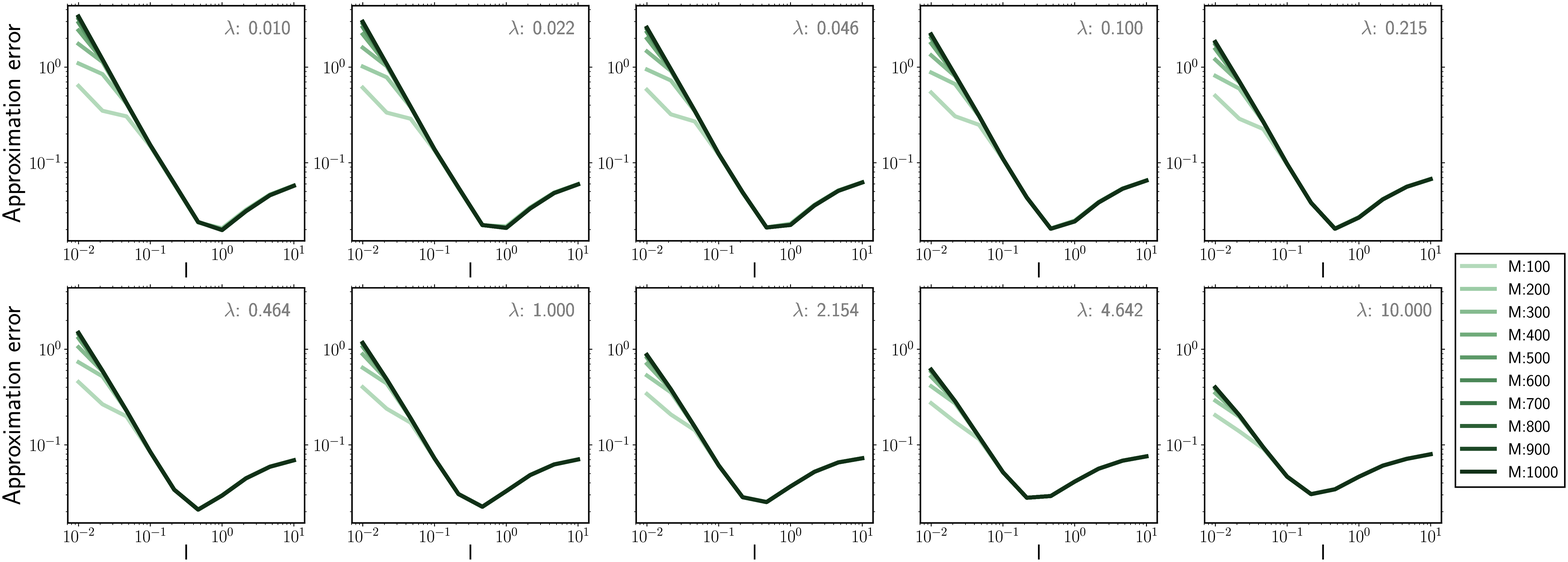}
  \caption{\textbf{Approximation error for increasing kernel length scale $l$ for different regularisation parameter values $\lambda$ and inducing point number $M$. }   
  }
  \label{fig:estimator1}
\end{figure}

\begin{figure}[htbp]
  \centering
  \includegraphics[width=1.1\textwidth]{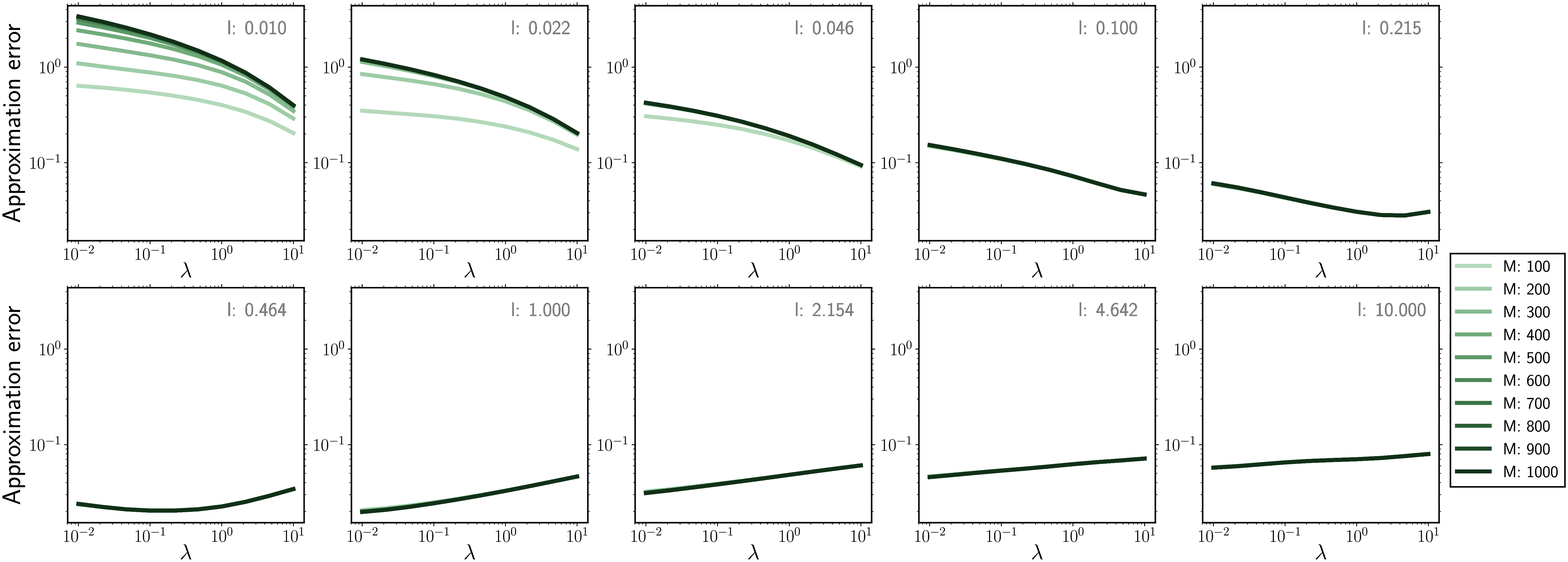}
  \caption{\textbf{Approximation error for increasing regularisation parameter value $\lambda$ for different kernel length scale $l$ and inducing point number $M$. }   
  }
  \label{fig:estimator2}
\end{figure}

By systematically varying the regularisation parameter $\lambda$, the kernel length scale $l$, and the inducing point number $M$ we observed the following:

\begin{itemize}
    \item[-] The hyperparameter that strongly influences the approximation accuracy is the kernel length scale $l$ (Fig.~\ref{fig:estimator1}).
    \item[-] Underestimation of kernel length scale $l$ has stronger impact on approximation accuracy, than overestimation (Fig.~\ref{fig:estimator1}).
    \item[-] For increasing regularisation parameter value $\lambda$, underestimation of $l$ has less impact on the approximation accuracy (Fig.~\ref{fig:estimator1} and Fig.~\ref{fig:estimator2}).
    \item[-] For overestimation of the kernel length scale $l$, regularisation parameter $\lambda$ and inducing point number $M$ have nearly no effect on the resulting approximation error (Fig.~\ref{fig:estimator1}).
    \item[-] For underestimation of kernel length scale $l$, increasing the number of inducing points $M$ in the estimator results in larger approximation errors (Fig.~\ref{fig:estimator2} (upper left)).
\end{itemize}

\section{Required number of particles for accurate Fokker--Planck solutions}

To compare the computational demands of the deterministic and stochastic particle systems we determined the required particle number each system needed to attain a specified accuracy to ground truth transient solutions. In particular, for a two dimensional Ornstein--Uhlenbeck process we identified the minimal number of particles $N^*_{KL}$ both systems required to achieve a certain time averaged Kullback--Leibler distance to ground truth transient solutions, $\langle \mathrm{KL}\left( P^A_{t}, P^N_{t}\right) \rangle_t$. 
As already indicated in the previous sections, the stochastic system required considerably larger particle number to achieve the same time averaged KL distances to ground truth when compared to our proposed framework. In fact, for the entire range of examined KL distances, our method consistently required at least one order of magnitude less particles compared to the its stochastic counterpart.   

\begin{figure}[htbp]
  \centering
  \includegraphics[width=0.61\textwidth]{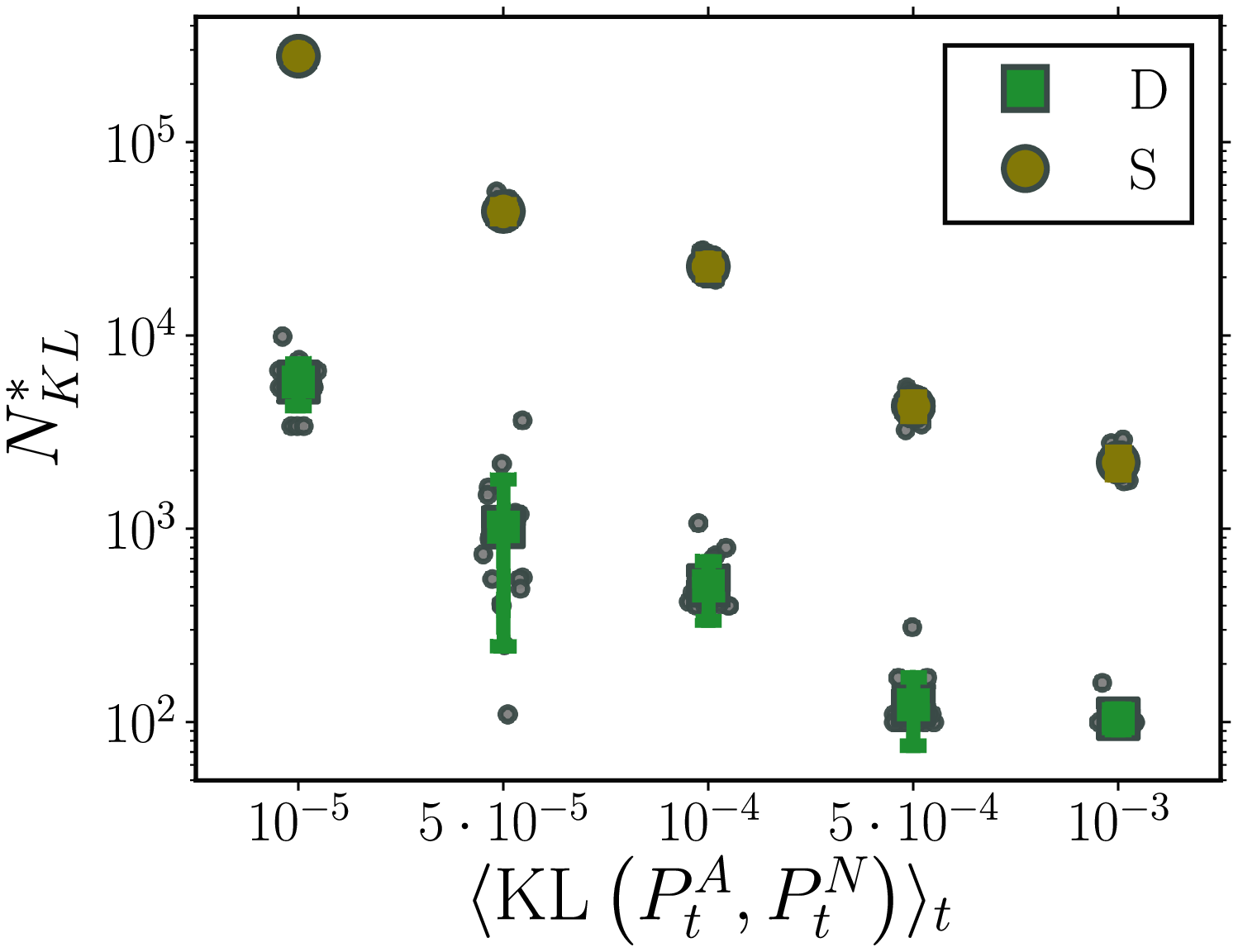}
  \caption{\textbf{Required particle number, $N^*_{KL}$, to attain time averaged Kullback--Leibler divergence to ground truth, $\langle \mathrm{KL}\left( P^A_{t}, P^N_{t}\right) \rangle_t$, for deterministic (green) and stochastic (brown) particle systems for a two dimensional Ornstein-Uhlenbeck process. } Markers indicate mean required particle number, while error bars denote one standard deviation over $20$ independent realisations. Grey circles indicate required particle number for each individual realisation. Deterministic particle system consistently required at least one order of magnitude less particles compared to its stochastic counterpart. (Further parameters values: regularisation constant $\lambda=0.001$, inducing point number $M=100$, and RBF kernel length scale $l$ estimated at every time point as two times the standard deviation of the state vector. Inducing point locations were selected randomly at each time step from a uniform distribution spanning the state space volume covered by the state vector.)  
  }
  \label{fig:complexity}
\end{figure}

\section{Algorithm for simulating deterministic particle system} \label{ap:algorithm}

Here we provide the algorithm for simulating deterministic particle trajectories according to our proposed framework. In the comments, we denote the computational complexity of each operation in the gradient--log--density estimation in terms of big-$\mathcal{O}$ notation. Since the inducing point number $M$ employed in the gradient--log--density estimation is considerably smaller than sample number $N$, i.e. $M\ll N$, the overall computational complexity of a \emph{single} gradient-log-density evaluation amounts to $\mathcal{O}\left( N\,M^2 \right)$.

\begin{algorithm}[H]
\SetAlgoLined
\DontPrintSemicolon
\KwIn{
    $X$: $N\times D$ state vector\\
    \hspace*{35pt}$Z$: $M\times D$ inducing points vector\\
    \hspace*{35pt}$d$: dimension for gradient \\
    \hspace*{35pt}$l$: RBF Kernel length scale\\
 }

\KwOut{$G$: $N\times 1$ vector for gradient-log-density at each position $X$ in $d$ dimension }
\BlankLine

$K^{xz} \longleftarrow K(X,Z;l)$    \tcp*{$N\times M$, $\mathcal{O}\left( N\,M \right)$}
$K^{zz} \longleftarrow K(Z,Z;l)$ \tcp*{$M\times M$, $\mathcal{O}\left( M^2 \right)$}
$I\_K^{zz} \longleftarrow \left(K^{zz} + 10^{-3}\,I\right)^{-1}$ \tcp*{$M\times M$, $\mathcal{O}\left( M^3 \right)$}
$grad\_K \longleftarrow \nabla_{X^{(d)}} K(X,Z;l) $ \tcp*{$N\times M$, $\mathcal{O}\left( N\,M \right)$}
$sgrad\_K \longleftarrow \sum\limits_{X_i} grad\_K $ \tcp*{$1\times M$}
$G \longleftarrow K^{xz} \, \left( \lambda \, I + I\_K^{zz} \,\left({K^{xz}}\right)^{\intercal}\,K^{xz} + 10^{-3}\,I  \right)^{-1}\, I\_K^{zz}\,   sgrad\_K^{\intercal}$   \tcp*{$N\times 1$}
\tcp*{, $\mathcal{O}\left( N\,M^2 \right) + \mathcal{O}\left( M^3 \right) $}
\caption{Gradient Log Density Estimator }
\end{algorithm}

\newpage

\begin{algorithm}[H]
\SetAlgoLined
\DontPrintSemicolon
\SetKwFunction{GradientLogDensityEstimation}{GradientLogDensityEstimation}
\SetKwFunction{Uniform}{Uniform}
\SetKwFunction{CreateRegularGrid}{CreateRegularGrid}
\SetKwFunction{mymin}{min}
\SetKwFunction{mymax}{max}
\KwIn{
     $x_0$: $1\times D$ initial condition\\
     \hspace*{35pt}$s_0$: variance of initial condition\\
     \hspace*{35pt}$N$: particle number\\
     \hspace*{35pt}$M$: inducing point number\\
     \hspace*{35pt}$T$: duration of simulation\\
     \hspace*{35pt}$dt$: integration time step\\
     \hspace*{35pt}$f(\cdot)$: drift function\\
     \hspace*{35pt}$D(\cdot)$: diffusion function\\
     \hspace*{35pt}$l\_0$: $1\times D$ Kernel length scale or False \tcp*{FALSE for adaptive length scale selection}
     \nonl\hspace*{35pt}$random\_M$: boolean variable \tcp*{TRUE for selecting inducing points from random uniform} \tcp*{distribution; FALSE for arranging them on a regular grid}
  }

\KwOut{$\{X_{t}\}^{{T}/{dt}}_{t=0}$: $N\times D \times \left \lceil{\frac{T}{dt}}\right \rceil  $ particle trajectories}
\BlankLine
Initialization: $X_0 \leftarrow$ Draw $N$ samples from Gaussian $\mathcal{N}\left( x_0,s_0  \right)$

\For{$t\leftarrow 1$ \KwTo $\frac{T}{dt}$}{

\eIf{$random\_M$\, is TRUE}{ \tcp*{Select inducing points}
    \For{$d\leftarrow 1$ \KwTo $D$}{
        $Z^{\left(d\right)}$ $\longleftarrow$  Draw $M$ samples from \Uniform{\mymin{$X^{\left(d\right)}_{t-1}$}, \mymax{$X^{\left(d\right)}_{t-1}$}}
    }
}{
 \For{$d\leftarrow 1$ \KwTo $D$}{
        $Z^{\left(d\right)}$ $\longleftarrow$  \CreateRegularGrid{\mymin{$X^{\left(d\right)}_{t-1}$}, \mymax{$X^{\left(d\right)}_{t-1}$}, M}
    }
}

\eIf{$l\_0$\, is FALSE}{ \tcp*{Set length scale}
$l$ $\longleftarrow$ $2\, std(X_{t-1})$
}{ $l$ $\longleftarrow$ $l\_0$}

\For{$d\leftarrow 1$ \KwTo $D$}{
    $G^{\left(d\right)}$ $\longleftarrow$ \GradientLogDensityEstimation{$X_{t-1}$,Z,d,l}
}
$X_{t}$ $\longleftarrow$ $X_{t-1} +  \left( f(X_{t-1}) -  \frac{1}{2} D(X_{t-1}) \circ G  -\frac{1}{2} \nabla D(X_{t-1})  \right) \, dt$
}
\caption{Deterministic Particle Simulation}
\end{algorithm}

\end{document}